**The Arpu Kuilpu Meteorite: In-depth characterization of an H5 chondrite delivered from a Jupiter Family Comet orbit.**

Seamus L. Anderson*[1], Gretchen K. Benedix[1,2], Belinda Godel[3], Romain M. L. Alosius[4], Daniela Krietsch[4], Henner Busemann[4], Colin Maden[4], Jon M. Friedrich[5,6], Lara R. McMonigal[5], Kees C. Welten[7], Marc W. Caffee[8,9], Robert J. Macke[10], Seán Cadogan[11], Dominic H. Ryan[12] Fred Jourdan[13], Celia Mayers[13], Matthias Laubenstein[14], Richard C. Greenwood[15], Malcom P. Roberts[16], Hadrien A. R. Devillepoix[1,17], Eleanor K. Sansom[1,17], Martin C. Towner[1], Martin Cupák[1,17,18], Philip A. Bland[1], Lucy V. Forman[1], John H. Fairweather[1], Ashley F. Rogers[1], Nicholas E. Timms[1]

[1]*Space Science and Technology Centre, School of Earth and Planetary Sciences, Curtin University, GPO Box U1987, Perth, WA, Australia.*
[2]*Dept. Earth and Planetary Sciences, Western Australian Museum, Locked Bag 49, Welshpool, WA, Australia.*
[3]*CSIRO Mineral Resources, Kensington, WA, Australia.*
[4]*Institute of Geochemistry and Petrology, ETH Zurich, CH-8092 Zurich, Switzerland.*
[5]*Dept. of Chemistry, Fordham University, 441 East Fordham Rd. Bronx, NY 10458 USA.*
[6]*Dept. of Earth and Planetary Sciences, American Museum of Natural History, 79th Street at Central Park West, New York, NY 10024, USA.*
[7]*Space Sciences Laboratory, University of California, Berkeley, 7 Gauss Way, Berkeley, CA 94720, USA.*
[8]*Dept. Physics and Astronomy, Purdue University, West Lafayette, IN 47907, USA.*
[9]*Dept. of Earth, Atmospheric and Planetary Sciences, West Lafayette, IN 47907, USA.*
[10]*Vatican Observatory V-00120 Vatican City-State.*
[11]*School of Science, UNSW Canberra, BC2610, Canberra, ACT, Australia.*
[12]*Dept. of Physics, McGill University, 3600 rue University, Montréal, QC, Canada.*
[13]*Western Australian Argon Isotope Facility, John de Laeter Centre, Curtin University, GPO Box U1987, Perth, WA 3845, Australia.*
[14]*National Institute of Nuclear Physics, National Laboratory at Gran Sasso, Via G. Acitelli, 22, 67100 Assergi L'Aquila, Italy.*
[15]*PSSRI, Open University, Milton Keynes, UK MK7 6AA.*
[16]*Centre for Microscopy, Characterisation and Analysis: University of Western Australia, 35 Stirling Hwy, Crawley WA 6009, Australia.*
[17]*International Centre for Radio Astronomy Research, Curtin University, GPO Box U1987, Perth, WA, Australia.*
[18]*Curtin Institute for Computation, Curtin University, GPO Box U1987, Perth, WA, Australia.*

*corresponding author: seamus.l.anderson@gmail.com

**ABSTRACT**

Over the Nullarbor Plain in South Australia, the Desert Fireball Network detected a fireball on the night of 1 June 2019 (7:30 pm local time), and six weeks later recovered a single meteorite (42 g) named Arpu Kuilpu. This meteorite was then distributed to a consortium of collaborating institutions to be measured and analyzed by a number of methodologies including: SEM-EDS, EPMA, ICP-MS, gamma-ray spectrometry, ideal gas pycnometry, magnetic susceptibility measurement, μCT, optical microscopy, and accelerator and noble gas mass spectrometry techniques. These analyses revealed that Arpu Kuilpu is an unbrecciated H5 ordinary chondrite, with minimal weathering (W0-1) and minimal shock (S2). The olivine and pyroxene mineral compositions (in mol%) are Fa: 19.2 ± 0.2, and Fs: 16.8 ± 0.2, further supporting the H5 type and class. The measured oxygen isotopes are also consistent with an H chondrite ($\delta^{17}$O‰ = 2.904 ± 0.177; $\delta^{18}$O‰ = 4.163 ± 0.336; $\Delta^{17}$O‰ = 0.740 ± 0.002). Ideal gas pycnometry measured bulk and grain densities of 3.66 ± 0.02 and 3.77 ± 0.02 g cm$^{-3}$, respectively, yielding a porosity of 3.0 % ± 0.7. The magnetic susceptibility of this meteorite is log χ = 5.16 ± 0.08. The most recent impact-related heating event experienced by Arpu Kuilpu was measured by $^{40}$Ar/$^{39}$Ar chronology to be 4467 ± 16 Ma, while the cosmic ray exposure age is estimated to be between 6-8 Ma. The noble gas isotopes, radionuclides, and fireball observations all indicate that Arpu Kuilpu's meteoroid was quite small (maximum radius of 10 cm, though more likely between 1-5 cm). Although this meteorite is a rather ordinary ordinary chondrite, its prior orbit resembled that of a Jupiter Family Comet (JFC) further lending support to the assertion that many cm- to m-sized objects on JFC orbits are asteroidal rather than cometary in origin.

**INTRODUCTION**

Although meteorites continuously fall to Earth, most lack any spatial context with regards to their orbital origin. Fireball observatory networks (Devillepoix et al. 2020; Oberst et al. 1998; Trigo-Rodríguez et al. 2006; and others) such as the Desert Fireball Network (DFN; Bland et al. 2012) are uniquely enabled to observe fireballs in order to calculate both their prior orbit, and the approximate landing site of the resulting meteorite. To date, 54 meteorites with associated orbits have been recovered and analyzed (Meier, 2017; see: https://www.meteoriteorbits.info/) in this effort to better understand how cm- to m-sized material is transferred from their asteroidal source bodies across the inner solar system. One such meteorite, Arpu Kuilpu (H5; Gattacecca et al. 2022), was observed by the DFN to fall in South Australia on 1$^{st}$ June 2019 at approximately 7:30 pm local time, near the Hughes train siding close to the border with Western Australia (Shober et al. 2022). This meteorite was a particularly high-priority target for recovery as it impacted Earth from a Jupiter family Comet (JFC) orbit (a=2.75 ± 0.03 AU; e=0.671 ± 0.003; i=2.03 ± 0.01 deg; T$_j$=2.97 ± 0.02 (Shober et al. 2022)), a rare origin for most orbitally constrained meteorites. A meteorite searching team was sent to the fall site six weeks later, recovering a single rock (42 g) on the 2$^{nd}$ day. The name Arpu Kuilpu (Are-puh Koo-ill-puh) was provided by the traditional custodians of the area, the Maralinga people.

While the work by Shober et al. (2022) details the astronomical observations (see their Fig. 1 and Fig. 3) and modelling of the fireball event as well as the recovery and classification of the meteorite as an H5 chondrite, this work focuses primarily on the physical, chemical, geological, and chronological aspects of the meteorite itself to provide a combined history for the formation of Arpu Kuilpu within its asteroidal parent body in the early solar system, as well as its more recent ejection as a meteoroid from its contemporary source body, and how it may relate to other orbital meteorites.

**METHODS**

We assembled a consortium (**Table 1**) of scientists from nearly a dozen institutions to perform a wide array of analyses to properly characterize many aspects of this meteorite.

**Table 1**. The Arpu Kuilpu Meteorite Consortium

| Science Goal | Method | Institution | Sample |
|---|---|---|---|
| Short-Lived Cosmogenic Radionuclides | HPGe* Gamma-Ray Spectrometry | INFN* at Gran Sasso | Half rock |
| Meteorite Type | SEM-EDS* | Curtin Uni. JdLC* | Epoxy-mounted sample (Polished and Carbon Coated) |
| Meteorite Petrologic Type | Electron Probe Micro Analysis (EPMA) | UWA*, CMCA | Epoxy-mounted sample (Polished and Carbon Coated) |
| Shock State | Optical Microscopy | Curtin Uni. | Polished Thin Section |
| Density, Shock State | X-ray Computed Tomography (μCT) | CSIRO* and AMNH* | Whole rock and 430.3 mg Aliquot |
| Oxygen Isotope Content | Infrared-Assisted Fluorination | The Open Uni. | 100 mg Aliquot |
| Noble Gas Content (CRE* Age, Meteoroid Size, Gas Retention Age) | Custom 'Albatros' Mass Spectrometer | ETH* Zurich | 29.2, 27.1 mg Aliquots |
| Bulk Elemental Composition | ICP-MS* | Fordham Uni. | 103.3, 107.8, 128.1 mg Aliquots |
| CRE* Age | AMS* and ICP-OES* | UC Berkeley, Purdue Uni. | 240 mg Aliquot |
| Terrestrial Weathering State | Mössbauer Spectroscopy | McGill Uni. | 500 mg Aliquot |
| Impact or Thermal Resetting Age | $^{40}Ar/^{39}Ar$ Dating | Curtin Uni. JdLC* | Pyroxene and plagioclase aliquots |
| Density and Porosity | Ideal Gas Pycnometry and Laser Scanning | Vatican Observatory | 14.64 g Piece |
| Magnetic Susceptibility | Magnetic Susceptibility Meter | Vatican Observatory | |

*SEM-EDS: Scanning Electron Microscope-Energy Dispersive X-ray Spectroscopy; ICP-MS: Inductively Coupled Plasma Mass Spectrometry; AMS: Accelerator Mass Spectrometry; ICP-OES: Inductively Coupled Plasma Optical Emission Spectroscopy; HPGe: HyperPure Germanium; CRE: Cosmic Ray Exposure; AMNH: American Museum of Natural History; INFN: Italian National Institute for Nuclear Physics; JdLC: John de Laeter Centre; UWA: University of Western Australia; CSIRO: Commonwealth Scientific and Industrial Research Organisation; ETH: Swiss Federal Institute of Technology; CMCA: Centre for Microscopy, Characterization and Analysis.

*HPGe Gamma-Ray Spectrometry (Short-lived Radionuclides)*
Cosmic-ray-produced (cosmogenic) radionuclide concentrations were analyzed by means of non-destructive high purity germanium (HPGe) gamma-ray spectrometry, and the counting efficiencies have been calculated using thoroughly tested Monte Carlo codes. One specimen of Arpu Kuilpu was measured in the underground laboratories at the Laboratori Nazionali del Gran Sasso (LNGS) (Arpesella, 1996; Laubenstein, 2017) for 40.63 days (160 days after the fall date of 1 June 2019).

*Micro X-Ray Computed Tomography (µCT; Bulk Density and Petrofabric Analysis)*

The entire 42.24 g meteorite was imaged in 3D via X-ray computed tomography at CSIRO Mineral Resources (Kensington, WA) using a Zeiss Versa 520 3D X-ray microscope. The instrument settings were 120 kV and 10 W, producing an 11 µm voxel$^{-1}$ size (3D pixel). Additionally, a 430.3 mg sub-sample was imaged at the AMNH at a resolution of 12.8 µm voxel$^{-1}$ edge using polychromatic X-rays with the GE phoenix v|tome|x s µCT system operating a tungsten X-ray tube at 110 kV and 13.2 W. The AMNH µCT data were intended to be used to verify the homogeneity of the sample prior to further sub-sampling for elemental analysis; however, we examined the petrofabric texture of this sub-sample too.

To quantitatively examine the petrofabric results, we used the same techniques as described in Friedrich et al. (2008; 2017). In short, we digitally separated the high-density Fe-Ni metal grains within the µCT volume using the program Blob3D (Ketcham, 2005). Best fit ellipsoids were constructed around the largest 5000 metal grains; next, the intersection of a line passing through the long axis of the best fit ellipsoid and a hemisphere enclosing the ellipsoid are plotted on a lower hemisphere equal area stereographic projection. Statistical descriptors of the fabric tensor eigenvalues (Woodcock, 1977; Woodcock and Naylor, 1983; Jelinek, 1981) generated from the collective metal grain projection data and examination of the individual best fit ellipsoid shapes using ratios of the ellipsoid long, intermediate, and short axes (Zingg, 1935; Blott and Pye, 2008) allows for a quantitative and thorough examination of the 3D petrofabric defined by metal grain shape preferred orientations within the chondrite.

*Scanning Electron Microscopy Energy Dispersive Spectroscopy (SEM-EDS; Meteorite Type via Mineral Phase Abundance)*

Using the TESCAN TIMA Scanning Electron Microscope (SEM), located at the John de Laeter Centre at Curtin University, we obtained element maps of our 25 mm-diameter epoxy-mounted sample via energy dispersive spectroscopy (EDS), which used four PulsTor Energy, silicon drift, Peltier cooled detectors to collect data. Prior to this, the sample mount was polished and carbon-coated to mitigate surface charging. TIMA maps were collected using a 25 kV acceleration voltage, 6.0 nA beam current, 15 mm working distance, and 90 nm spot size. Seven element maps (Al, Ca, Fe, Cr, Mg, Si, S) were combined to create a mineral phase map using a python script (Anderson, 2023).

*Electron Probe MicroAnalysis (EPMA; Petrologic Type via Mineral Compositions)*

The major element compositions of olivine, orthopyroxene, and chromite in Arpu Kuilpu were measured using the JEOL 8530F Electron Probe MicroAnalyser (EPMA) at the Centre for Microscopy and Microanalysis, University of Western Australia. On the same polished sample mount used for SEM-EDS, we selected 20 olivine, 20 orthopyroxene, and 11 chromite grains for elemental measurements using a beam current of 20 nA and an accelerating voltage of 15 kV. We also collected calibration measurements of our mineral standards: wollastonite (Si and Ca), periclase (Mg), magnetite (Fe), manganese metal (Mn), corundum (Al), jadeite (Na), and Durango apatite (P).

*Optical Microscopy*

A polished thin section of Arpu Kuilpu was imaged in plane-polarized and cross-polarized light with x5 magnification using a Zeiss Axio Imager 503 (color), utilizing its accompanying software for automated stitching of a mosaic. A Leica microscope with a rotating stage, in transmitted light, was used to identify shock features in the same thin section.

### *Mössbauer Spectroscopy (Weathering State)*

We employed Mössbauer spectroscopy, a non-destructive and non-altering technique (apart from crushing into a powder), to quantify the abundances of $Fe^{2+}$ and $Fe^{3+}$ in the meteorite sample (Dyar et al., 2006). It can readily distinguish between magnetic and non-magnetic minerals, and since it is a bulk transmission measurement, it is sensitive to all iron-bearing components in the sample. Mössbauer spectroscopy is often used to characterize the weathering of minerals (Murrad and Cashion, 2004), and was an instrument on the Martian Spirit and Opportunity rovers (Morris et al., 2004).

The $^{57}$Fe Mössbauer spectrum of a 0.1 g aliquot of Arpu Kuilpu was measured at McGill University, at room temperature in transmission mode using a sinusoidal drive velocity. The velocity scale was calibrated with a standard α-Fe foil and all isomer shifts are quoted relative to the α-Fe calibration spectrum. The spectrum of Arpu Kuilpu was fitted using three paramagnetic doublets and two magnetic sextets.

### *Inductively Coupled Plasma Mass Spectrometry (ICP-MS; Bulk Major and Trace Element Geochemistry)*

We performed elemental analysis on three chips of Arpu Kuilpu (103.3, 107.8, 128.1 mg) at the Chemistry Department of Fordham University. We first ground each chip in an agate mortar and pestle. To dissolve the powders, we used a combination of HF and $HNO_3$ in a high pressure CEM Mars 5 microwave digestion system, dried the resulting mixture to incipient dryness on a hotplate, treated the residue with $HClO_4$, and again brought it to incipient dryness to dissolve the samples. In addition to the Arpu Kuilpu samples, a procedural blank and the Allende Standard Reference Meteorite were taken up in 1% $HNO_3$ and analyzed with a ThermoElemental X-Series II ICPMS with the methods outlined in Friedrich et al. (2003) for trace elements, and Wolf et al. (2012) for major and minor elements. For examination of the data, we compare them with Orgueil CI chondrite values from Friedrich et al. (2002) for trace elements and the compiled Orgueil results of Anders and Grevesse (1989) for major and minor elements.

### *Oxygen Isotopes*

The O isotope content of Arpu Kuilpu was measured at the Open University using an infrared laser-assisted fluorination system (Miller et al. 1999; Greenwood et al. 2017). The analyses were performed on 2 mg aliquots, sampled from ~100 mg of homogenized meteorite powder, which was prepared by crushing fusion-crust-free pieces of the meteorite. Oxygen was collected from the sample by heating it in the presence of $BrF_5$, then passing the resultant gas through two cryogenic nitrogen traps and a heated bed of KBr, finally being measured by a MAT 253 dual inlet mass spectrometer. The 1σ precision obtained for $δ^{17}O$ and $δ^{18}O$ was 0.08 ‰ and 0.04 ‰, respectively.

### *Laser Scanning and Ideal Gas Pycnometry (Density and Porosity)*

A 14.64 g piece of Arpu Kuilpu was used for the determination of bulk and grain density, as well as porosity. Bulk volume was measured with NextEngine model 2020i ScannerHDPro laser scanner with a rotating platform (Macke et al. 2015). Scans were performed at 16k dots per inch (dpi). For each of the three possible orientations of the meteorite, 16 scans were performed, with the platform rotated 1/16 of a complete circle between each scan. Scans were trimmed of artifacts and aligned to construct a "watertight" computer model of the specimen, from which we calculated the bulk volume. Grain volume was measured via ideal gas pycnometry with a Quantachrome Ultrapycnometer 1000, using gaseous nitrogen. The measurement procedure was run 15 times, of which the average of the last 6 were taken as the grain volume of the specimen, consistent with the technique described in Macke (2010). Porosity (P) was calculated from the bulk and grain densities (ρ): $P = 1 - ρ_{bulk}/ρ_{grain}$

*Magnetic Susceptibility*

Using the same 14.64 g piece analyzed for density and porosity (see above), the magnetic susceptibility of Arpu Kuilpu was measured using a ZH-instruments SM30 magnetic susceptibility meter. Volumetric corrections were applied according to the work of Gattacceca et al. (2004) with additional shape correction according to Macke (2010). In the context of this study, the Magnetic susceptibility of an ordinary chondrite, combined with its bulk and grain densities can provide a qualitative measure of the meteorite's class and weathering (Consolmagno et al. 2006).

*Cosmogenic Radionuclides (Cosmic Ray Exposure)*

We prepared samples by using a ~0.24 g chip of Arpu Kuilpu for analysis of the cosmogenic radionuclides $^{10}$Be (half-life=1.36 × 10$^6$ yr), $^{26}$Al (7.05 × 10$^5$ yr) and $^{36}$Cl (3.01 × 10$^5$ yr), following the procedures described in Welten et al. (2011). We crushed the sample in an agate mortar and separated the magnetic (metal) from the non-magnetic (stone) fraction. The magnetic fraction was purified by ultrasonic agitation in 0.2N HCl to remove attached troilite. After rinsing the metal four times with MilliQ water and once with ethanol, we obtained 33 mg of relatively clean metal, corresponding to 13.9 wt% bulk metal. The metal fraction was further purified by ultrasonic agitation in concentrated HF for 15 min to dissolve attached silicates. We dissolved 32 mg of the purified metal in ~10% HNO$_3$ along with a carrier solution containing 0.90 mg Be, 1.01 mg Al, and 2.99 mg Cl. After dissolution, we took a small aliquot (~5.5%) of the dissolved sample for chemical analysis and used the remaining solution for cosmogenic radionuclide analysis. We also dissolved 96 mg of the stone fraction of Arpu Kuilpu, along with 3.19 mg of Be carrier and 2.77 mg of Cl carrier, in concentrated HF/HNO$_3$ by heating the mixture for 24 h inside a Parr Teflon digestion bomb at 125 °C. After cooling the sample to room temperature, we separated the Cl fraction as AgCl, and removed Si in the solution by repeatedly fuming the sample to dryness with HClO$_4$. The residue was dissolved in dilute HCl and a small aliquot (3.4%) was taken for chemical analysis. After adding ~5.0 mg of additional Al carrier to the remaining solution, we separated the Be and Al fraction for radionuclide analysis.

After separating and purifying the Be, Al and Cl fractions, the $^{10}$Be/Be, $^{26}$Al/Al and $^{36}$Cl/Cl ratios of the samples, blanks and standards were measured by Accelerator Mass Spectrometry (AMS) at Purdue University's PRIME Lab (Sharma et al. 2000). The measured ratios were normalized to those of well-known AMS standards (Nishiizumi 2004; Nishiizumi et al. 2007; Sharma et al. 1990) and converted to concentrations in disintegrations per minute per kg (dpm/kg).

For the chemical analysis, we made two consecutive dilutions of the aliquots of the dissolved metal and stone fractions for characterization by ICP-OES. All dilutions were measured on a Thermo Fisher iCAP 6300 duo instrument.

*Noble Gas Composition (Cosmic Ray Exposure, Meteoroid size and Sample Shielding, Gas Retention Age)*

Measurements of all stable noble gas isotopes (He, Ne, Ar, Kr, Xe) were performed on two aliquots (29.267 ± 0.015 mg (AKL) and 27.178 ± 0.014 mg (AKS)) of Arpu Kuilpu. The samples were first wrapped in aluminum foil and heated at 110°C in ultra-high vacuum for several days to remove adsorbed atmospheric gases. The measurements were carried out on the in-house-built noble gas mass spectrometer "Albatros" at ETH Zürich, using the procedures described in detail in Riebe et al. (2017). Gas extraction was achieved by melting the samples in a Mo crucible at ~1700 °C for ~25 min. The blank corrections for both aliquots are <1.5% of the He and Ne isotope signals, and <2.5% for Ar. The corrections for all Kr isotopes were < 7% of the signals, and < 2% for all Xe isotopes.

We numerically separated the cosmogenic (cos) and trapped (tr) components using a two-component deconvolution between ($^{36}$Ar/$^{38}$Ar)$_{cos}$ (0.63-0.67; Wieler, 2002) and ($^{36}$Ar/$^{38}$Ar)$_{tr}$. The trapped component used for the deconvolution was delimited by the values for Q and air (5.32-5.34; Busemann et al. 2000; Nier, 1950) since no solar wind component was detected for He and Ne.

To constrain the CRE age from noble gas data, we calculated the production rates of cosmogenic $^{3}$He, $^{21}$Ne, and $^{38}$Ar based on the model for ordinary chondrites from Leya & Masarik (2009), which considers the pre-atmospheric size of the meteoroid, the depth of the sample within the meteoroid, and the bulk chemical composition of the sample. Where possible, we used the elemental concentrations reported by the ICP-MS portion of this study in these model calculations. For elements not measured in this study (via ICP-MS) we used data for H chondrites from the literature (Si and S from Alexander, 2019; C from Lodders and Fegley, 1998; while O was calculated for the sum of all concentrations to reach 100 wt%). Since the original observations of the fireball (Shober et al. 2022) suggest a pre-atmospheric meteoroid size of ~5 cm, we used a modified version (Wieler et al. 2016) of the Leya and Masarik (2009) model, which considers small (<7 cm) H chondrite meteoroids. Ratios of cosmogenic noble gases, specifically the cosmogenic $^{22}$Ne/$^{21}$Ne ratio can be used as a shielding indicator to then calculate the size of the original meteoroid and the meteorite's burial depth within it, which is required to determine production rates and CRE age. Consistent with the small pre-atmospheric meteoroid size assessed independently, only the modified model for small H chondritic meteoroids (Wieler et al. 2016) revealed matches between prediction and the measured high cosmogenic $^{22}$Ne/$^{21}$Ne ratios.

Gas retention ages for both aliquots were calculated separately using the U/Th-He and K-Ar chronometers with the U, Th, and K concentrations from the Gamma-Ray Spectrometry portion of this study. The $^{4}$He$_{rad}$ values for the U/Th-He chronometer were calculated by assuming that the $^{3}$He concentrations were cosmogenic, and applying a ($^{4}$He/$^{3}$He)$_{cos}$ ratio between 5.2 and 6.1 (Wieler, 2002). The $^{4}$He$_{cos}$ concentration was then subtracted from our measured $^{4}$He concentration (with $^{4}$He$_{tr}$ being negligible). The $^{40}$Ar$_{rad}$ values for the K-Ar chronometer were calculated with the deconvoluted $^{36}$Ar$_{tr}$, adopting a ratio ($^{40}$Ar/$^{36}$Ar)$_{tr}$ between 0 and 295.5 (covering Q and air composition; Busemann et al. 2000; Steiger and Jäger, 1977).

### *$^{40}$Ar/$^{39}$Ar Chronology (Thermal and Impact History)*

To determine the $^{40}$Ar/$^{39}$Ar chronology, we crushed a whole rock fragment from Arpu Kuilpu and selected nine pyroxene aliquots each including between 1 and 30 grains, and with each grain ranging in size from 150 to 350 µm in diameter. We also selected a single plagioclase grain . The samples were irradiated for 40 hours, then analyzed at the Western Australian Argon Isotope Facility at Curtin University using a ARGUS VI Mass Spectrometer, using a 10.4 µm CO$_2$ laser to affect step heating for 60 sec. The complete procedure is similar to the one described by Jourdan et al. (2020) for other meteorite samples.

**RESULTS**

### *SEM-EDS (Meteorite Class and Texture)*

The mineral map produced by SEM-EDS imaging is shown in **Figure 1**, and clearly reveals a texture consistent with an equilibrated ordinary chondrite, consisting of Fe-Ni metal (kamacite and taenite), chromite, troilite, plagioclase and Ca-phosphate grains, as well as olivine and pyroxene chondrules, all set within a recrystallized, mostly silicate matrix. There are recognizable chondrules, though they have fairly indistinct boundaries separating them from the matrix.

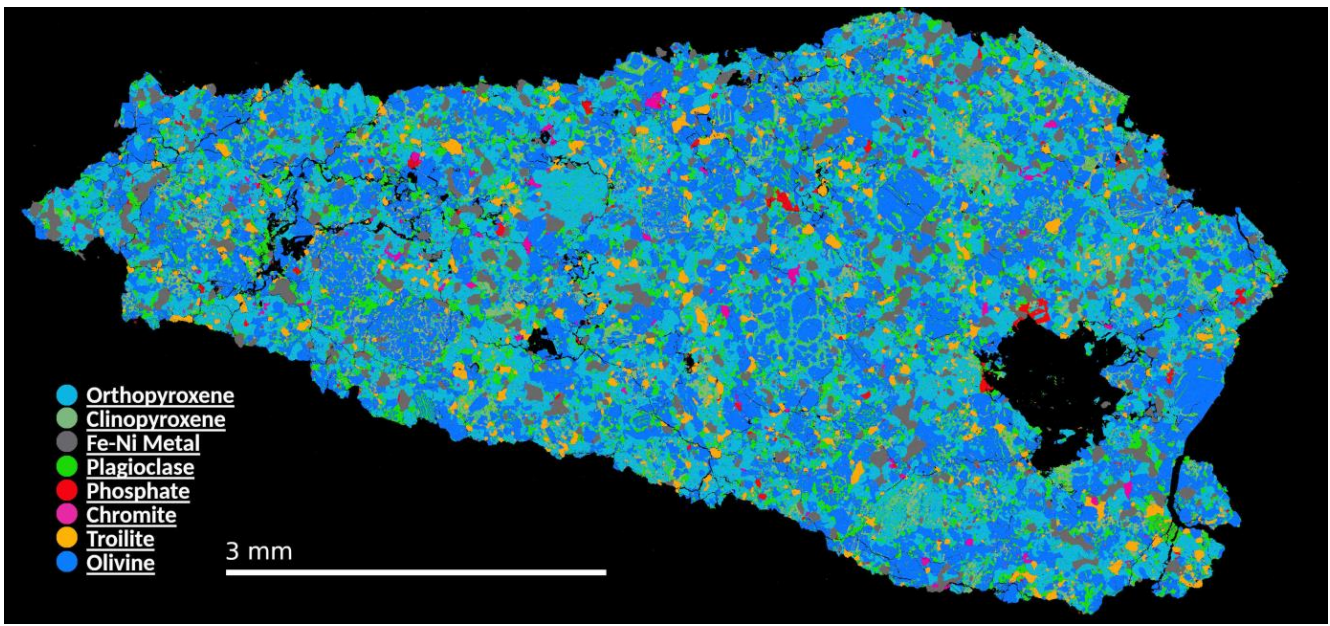

**Figure 1.** A false color mineral map of Arpu Kuilpu's epoxy-mounted thick section taken from SEM-EDS data (initially appearing as Fig. 9 in Shober et al. 2022), which clearly shows a texture typical of an equilibrated ordinary chondrite. The large void on the right is an artifact of the polishing step when a remnant chondrule was likely and unintentionally removed.

*EPMA (Electron Probe MicroAnalysis; Mineral Compositions)*

The full dataset from the EPMA is displayed in **Table 2.** The olivine compositions of Arpu Kuilpu average at fayalite = 19.2 ± 0.2 mol% (n=20), while the orthopyroxenes reside at ferrosilite = 16.8 ± 0.2 mol% and wollastonite = 1.4 ± 0.2 mol% (n=20). The chromite compositions are: Cr/(Cr + Al) = 85.5 ± 0.3, and Fe/(Fe + Mg) = 84.4 ± 1.2 (n=11). The olivine and pyroxene compositions reveal that Arpu Kuilpu is an H chondrite (Van Schmus and Wood, 1967), while the Wo value of 1.4 indicates a petrologic type 5 (Scott et al. 1986). All uncertainties listed above are 1σ.

**Table 2.** EPMA measurements of Arpu Kuilpu's Olivines, Orthopyroxenes, and Chromites. Oxides concentrations are displayed in wt%.

| SAMPLE | SiO$_2$ | MgO | Cr$_2$O$_3$ | FeO | MnO | CaO | Al$_2$O$_3$ | Na$_2$O | P$_2$O$_5$ | TOTAL | Fa (mol%) | |
|---|---|---|---|---|---|---|---|---|---|---|---|---|
| Olivine 1 | 39.37 | 42.59 | 0.02471 | 18.00 | 0.5465 | 0.02976 | 0.006649 | 0 | 0 | 100.6 | 19.17 | |
| Olivine 2 | 39.76 | 42.78 | 0 | 18.18 | 0.5472 | 0.03659 | 0 | 0 | 0.01659 | 101.3 | 19.25 | |
| Olivine 3 | 39.40 | 42.58 | 0 | 17.96 | 0.3904 | 0 | 0 | 0.008298 | 0.01331 | 100.4 | 19.13 | |
| Olivine 4 | 39.02 | 42.65 | 0.01216 | 17.98 | 0.4182 | 0.01036 | 0.01977 | 0 | 0 | 100.1 | 19.12 | |
| Olivine 5 | 39.68 | 43.07 | 0.006442 | 18.04 | 0.4237 | 0.03139 | 0 | 0.05316 | 0 | 101.3 | 19.02 | |
| Olivine 6 | 39.69 | 42.81 | 0.008865 | 18.32 | 0.4694 | 0.02026 | 0.00836 | 0.0336 | 0 | 101.4 | 19.36 | |
| Olivine 7 | 39.49 | 42.77 | 0.01327 | 18.19 | 0.5314 | 0 | 0.03929 | 0 | 0.0629 | 101.1 | 19.27 | |
| Olivine 8 | 39.38 | 42.71 | 0 | 18.24 | 0.4375 | 0.02325 | 0 | 0.00719 | 0.05583 | 100.8 | 19.33 | |
| Olivine 9 | 39.47 | 42.74 | 0.02494 | 17.79 | 0.5301 | 0.02125 | 0 | 0.04221 | 0 | 100.6 | 18.93 | |
| Olivine 10 | 39.33 | 42.65 | 0.01021 | 17.69 | 0.3821 | 0.004424 | 0 | 0.02753 | 0.02789 | 100.1 | 18.88 | |
| Olivine 11 | 39.35 | 42.34 | 0.01959 | 17.75 | 0.4737 | 0.02639 | 0.01736 | 0 | 0.002732 | 99.97 | 19.04 | |
| Olivine 12 | 39.79 | 42.92 | 0 | 17.78 | 0.4709 | 0 | 0 | 0 | 0 | 101.0 | 18.86 | |
| Olivine 13 | 39.58 | 42.86 | 0 | 18.23 | 0.4658 | 0.02285 | 0.01419 | 0 | 0.01676 | 101.2 | 19.26 | |
| Olivine 14 | 39.84 | 43.19 | 0.02413 | 18.10 | 0.4678 | 0.03081 | 0 | 0.004621 | 0.006275 | 101.7 | 19.03 | |
| Olivine 15 | 39.39 | 42.72 | 0.01905 | 18.25 | 0.5195 | 0.0193 | 0 | 0.01213 | 0.002416 | 100.9 | 19.34 | |
| Olivine 16 | 39.58 | 42.63 | 0.01839 | 18.25 | 0.4682 | 0.006519 | 0 | 0 | 0 | 101.0 | 19.37 | |
| Olivine 17 | 39.59 | 42.83 | 0.03658 | 18.16 | 0.5148 | 0.02594 | 0 | 0 | 0.0133 | 101.2 | 19.22 | |
| Olivine 18 | 39.52 | 42.71 | 0 | 18.29 | 0.4783 | 0.01196 | 0 | 0.01093 | 0.002497 | 101.0 | 19.37 | |
| Olivine 19 | 39.55 | 42.75 | 0.04754 | 18.12 | 0.5177 | 0.01065 | 0 | 0.007438 | 0 | 101.0 | 19.21 | |
| Olivine 20 | 39.38 | 42.76 | 0.006419 | 18.27 | 0.3563 | 0.002413 | 0 | 0.0637 | 0.03473 | 100.9 | 19.33 | |
| | | | | | | | | | | | Fs (mol%) | Wo (mol%) |
| Orthopyroxene 1 | 57.19 | 31.02 | 0.2421 | 11.02 | 0.5814 | 0.8352 | 0.1479 | 0.04213 | 0.07061 | 101.2 | 16.35 | 1.588 |
| Orthopyroxene 2 | 57.07 | 31.20 | 0.3319 | 11.28 | 0.5246 | 0.5783 | 0.1389 | 0.03082 | 0.005614 | 101.2 | 16.67 | 1.095 |
| Orthopyroxene 3 | 57.38 | 31.11 | 0.1699 | 11.60 | 0.4985 | 0.8453 | 0.1817 | 0.0193 | 0 | 101.8 | 17.02 | 1.590 |
| Orthopyroxene 4 | 57.01 | 30.72 | 0.1212 | 11.17 | 0.4074 | 0.7721 | 0.1094 | 0.0464 | 0 | 100.4 | 16.69 | 1.478 |
| Orthopyroxene 5 | 56.68 | 30.81 | 0.1051 | 11.42 | 0.4455 | 0.5648 | 0.1280 | 0.06928 | 0 | 100.2 | 17.02 | 1.079 |
| Orthopyroxene 6 | 56.78 | 30.88 | 0.1249 | 11.62 | 0.4816 | 0.8609 | 0.1725 | 0 | 0 | 100.9 | 17.15 | 1.628 |
| Orthopyroxene 7 | 57.32 | 31.16 | 0.06102 | 11.22 | 0.5290 | 0.5903 | 0.1248 | 0 | 0 | 101.0 | 16.62 | 1.120 |
| Orthopyroxene 8 | 57.16 | 30.97 | 0.1389 | 11.29 | 0.5858 | 0.5970 | 0.1896 | 0.01186 | 0 | 101.0 | 16.78 | 1.137 |
| Orthopyroxene 9 | 57.01 | 31.03 | 0.1176 | 11.32 | 0.3672 | 0.6166 | 0.1507 | 0 | 0 | 100.6 | 16.78 | 1.172 |
| Orthopyroxene 10 | 57.02 | 30.69 | 0.1346 | 11.47 | 0.6185 | 0.8632 | 0.1974 | 0 | 0 | 101.0 | 17.05 | 1.644 |
| Orthopyroxene 11 | 57.19 | 30.85 | 0.0927 | 11.50 | 0.4944 | 0.7872 | 0.1883 | 0.01477 | 0 | 101.1 | 17.04 | 1.494 |
| Orthopyroxene 12 | 57.08 | 30.80 | 0.1348 | 11.47 | 0.5524 | 0.9027 | 0.1616 | 0 | 0 | 101.1 | 16.98 | 1.713 |
| Orthopyroxene 13 | 57.26 | 31.08 | 0.1518 | 11.33 | 0.5272 | 0.7350 | 0.1577 | 0.02716 | 0 | 101.3 | 16.74 | 1.391 |
| Orthopyroxene 14 | 57.03 | 31.06 | 0.08956 | 11.34 | 0.2964 | 0.5215 | 0.1755 | 0 | 0.005946 | 100.5 | 16.83 | 0.9914 |
| Orthopyroxene 15 | 57.29 | 31.11 | 0.1410 | 11.70 | 0.5075 | 0.6951 | 0.1219 | 0.02795 | 0.0126 | 101.6 | 17.20 | 1.309 |
| Orthopyroxene 16 | 56.85 | 30.82 | 0.1416 | 11.39 | 0.3288 | 0.7306 | 0.1164 | 0.02718 | 0.02725 | 100.4 | 16.93 | 1.392 |
| Orthopyroxene 17 | 57.41 | 31.04 | 0.1250 | 11.42 | 0.5959 | 0.8248 | 0.1293 | 0 | 0 | 101.5 | 16.85 | 1.559 |
| Orthopyroxene 18 | 57.19 | 30.91 | 0.1288 | 11.11 | 0.4772 | 0.7625 | 0.1339 | 0.02558 | 0 | 100.7 | 16.53 | 1.454 |
| Orthopyroxene 19 | 57.10 | 30.82 | 0.1908 | 11.07 | 0.5066 | 0.8067 | 0.1410 | 0 | 0 | 100.6 | 16.51 | 1.542 |
| Orthopyroxene 20 | 57.10 | 30.67 | 0.1646 | 11.15 | 0.5058 | 0.8799 | 0.1500 | 0.01548 | 0.002003 | 100.6 | 16.66 | 1.684 |
| | | | | | | | | | | | Cr/(Cr+Al) | Fe/(Fe+Mg) |
| Chromite 1 | 0.007661 | 2.929 | 56.82 | 29.93 | 0.7726 | 0.01477 | 6.393 | 0.01192 | 0 | 96.87 | 85.64 | 85.14 |
| Chromite 2 | 0 | 2.959 | 56.14 | 29.86 | 0.9041 | 0 | 6.670 | 0.03961 | 0 | 96.57 | 84.95 | 84.99 |
| Chromite 3 | 0 | 3.362 | 56.38 | 29.43 | 0.7047 | 0.006257 | 6.516 | 0.05362 | 0 | 96.46 | 85.30 | 83.08 |
| Chromite 4 | 0 | 2.690 | 56.75 | 30.57 | 0.8618 | 0.04033 | 6.621 | 0 | 0 | 97.53 | 85.18 | 86.44 |
| Chromite 5 | 0.04861 | 3.383 | 57.36 | 28.85 | 0.8402 | 0.005345 | 6.641 | 0.01882 | 0 | 97.14 | 85.28 | 82.71 |
| Chromite 6 | 0 | 3.444 | 57.65 | 29.09 | 0.6289 | 0.02526 | 6.343 | 0.1256 | 0.04955 | 97.35 | 85.91 | 82.57 |
| Chromite 7 | 0 | 2.943 | 56.75 | 29.93 | 0.6614 | 0.03826 | 6.328 | 0 | 0.002324 | 96.66 | 85.75 | 85.09 |
| Chromite 8 | 0 | 2.921 | 56.80 | 29.80 | 0.7989 | 0 | 6.401 | 0 | 0 | 96.72 | 85.62 | 85.13 |
| Chromite 9 | 0.03422 | 3.077 | 56.41 | 30.14 | 0.6484 | 0.003506 | 6.390 | 0 | 0.005795 | 96.70 | 85.55 | 84.60 |
| Chromite 10 | 0 | 3.113 | 56.82 | 29.55 | 0.9405 | 0 | 6.295 | 0.1222 | 0.02445 | 96.87 | 85.83 | 84.19 |
| Chromite 11 | 0 | 3.129 | 56.76 | 29.90 | 0.5190 | 0.009928 | 6.273 | 0 | 0 | 96.59 | 85.86 | 84.28 |

*Optical Microscopy*

The optical mosaic photomicrographs of the thin section, taken using the Zeiss Axio are shown in **Figure 2.** Further manual investigation using a rotating stage revealed both sharp and undulose extinction features in olivine and plagioclase grains indicating this meteorite experienced very low shock-induced pressures or temperatures, equating to S1-S2 in the Stöffler et al. (2018) classification scheme. No opaque shock veins or mosaicism were observed, eliminating the possibility for localized S3 shock or higher.

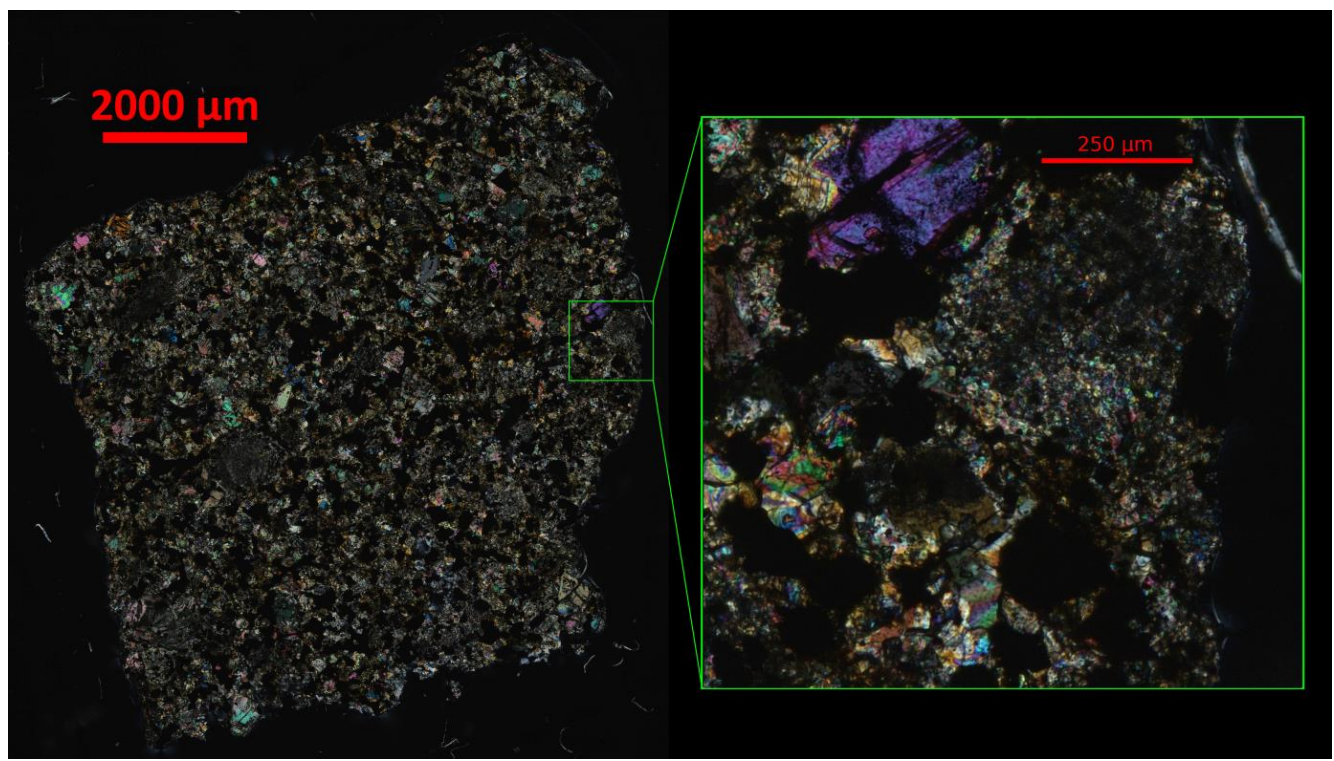

**Figure 2.** Cross-polarized image of our Arpu Kuilpu thin section, which clearly shows a texture consistent with an equilibrated H chondrite. Further investigation of this thin section under a rotating stage reveals undulose light extinction in olivine and plagioclase grains, indicating a low-pressure shock history (S 1-2; Stöffler et al. 2018).

*Mössbauer Spectroscopy (Weathering State)*

The spectrum of the Arpu Kuilpu meteorite sample shown in Figure 3 is dominated by a pair of peaks at ∼0 mm/s and +2.5 mm/s. This wide split doublet, offset to positive velocities, is characteristic of a ferrous (Fe2+) mineral. The distance between the lines is the quadrupole splitting (QS) and is associated with an electric field gradient at the iron site in the mineral, while the center of the doublet (arithmetic mean of the two peak positions) is the isomer shift (IS). Closer inspection of the two peaks reveals that each of the lines has a second, weaker line associated with it and these two weaker lines are due to a second Fe2+ mineral. The fitted values of the QS and IS for these components, given in Table 3, allow them to be identified as arising from Olivine and Pyroxene, with the former mineral being about twice as abundant as the latter (Dyar et al., 2006).

To the left and right of the primary doublets there are a series of weaker peaks. These are due to the presence of two magnetic minerals: Fe-Ni metal, and Troilite. As with Olivine and Pyroxene, the two magnetic minerals are identified using the fitted parameters (IS, QS and hyperfine field) shown in

Table 3. Finally, we observed a weak ferric component in the spectrum that accounted for ~2% of the total area and is attributed to paramagnetic nanoparticles. The results of the Mössbauer analysis show little terrestrial alteration, as any $Fe^{3+}$ signal in an ordinary chondrite (which is quite low for Arpu Kuilpu) would originate only from terrestrial alteration.

**Table 3.** The values of Isomer Shift (I.S.), and Quadrupole Splitter (Q.S). for subspectra #1 and #2 indicate $Fe^{2+}$ and are assigned to olivine and pyroxene, respectively. Subspectrum #3 indicates $Fe^{3+}$ paramagnetic nanoparticles (n-p-3+). Component #4 is a magnetic sextet with small I.S. and Q.S. values, and a magnetic Hyperfine Field (H.F.F.) of 33.59 T indicating Fe metal. Component #5 is a magnetic sextet whose I.S. and H.F.F. values indicate Troilite.

| Subspectrum No. | I.S. (mm/s) | Q.S. (mm/s) | H.F.F. (T) | Area (%) | Likely Component |
|---|---|---|---|---|---|
| 1 | 1.135 | 2.96 | - | 45.5 | Olivine |
| 2 | 1.132 | 2.094 | - | 21.6 | Pyroxene |
| 3 | 0.428 | 0.688 | - | 1.8 | n-p-3+ |
| 4 | 0.009 | -0.003 | 33.59 | 12.7 | Fe-Ni Metal |
| 5 | 0.752 | -0.147 | 31.35 | 18.3 | Troilite |

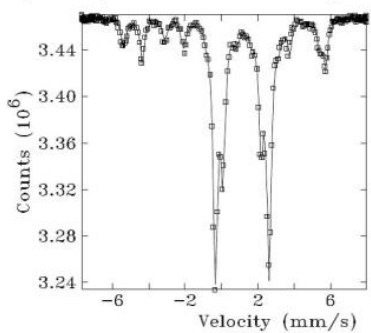
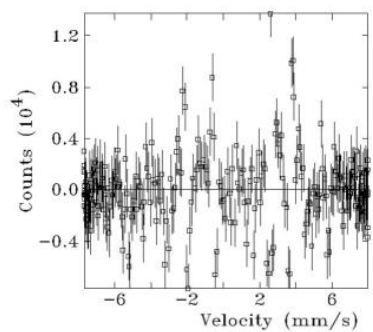

**Figure 3**. The Mössbauer spectrum of Arpu Kuilpu (top) and the fitting difference (bottom), note the dramatically different vertical scales.

*ICP-MS (Bulk Major and Trace Element Analysis)*

The 50 major, minor, and trace elements measured via ICP-MS are shown in **Table 4**. The higher inter-aliquot variability of the light REE may be due to one of the aliquots being very slightly enriched in phosphates or having an unrepresentative pyroxene/plagioclase ratio relative to the whole sample (Mason and Graham, 1970). The Fe, Ni, and Co abundances normalized to CI and Mg (**Figure 4**) of Arpu Kuilpu are more consistent with what would be expected for an H chondrite than the siderophile trace elements Re, Ir, Mo, Pt, and Pd which are all nearly 40% higher in abundance (normalized to CI and Mg) than the major elements Fe and Ni. Perhaps we are seeing a refractory siderophile nugget affecting enrichment of the refractory siderophile elements within our analyzed aliquots.

**Table 4**. Abundances of 50 elements in the Arpu Kuilpu chondrite. Errors, given as percent relative standard deviation (%RSD), are ≤14.0% for all elements except for the light Rare Earth Elements La, Ce, Pr, Nd, and Sm, where all had %RSD values between 16.5-17.6 based on the three replicate aliquots analyzed. Other more variable elements include Re (17.6%), Ir (25.0%), Sn (20.0%), Cu (21.2%), As (16.6%), Sb (37.9%).

| Element | Units | Value |
|---|---|---|
| Li | µg/g | 1.3 |
| Na | mg/g | 6.2 |
| Mg | mg/g | 135 |
| Al | mg/g | 10.9 |
| K | mg/g | 0.83 |
| Ca | mg/g | 11.9 |
| Sc | µg/g | 8.6 |
| Ti | µg/g | 71 |
| Mn | mg/g | 2.3 |
| Fe | mg/g | 276 |
| Co | µg/g | 870 |
| Ni | mg/g | 18.4 |
| Cu | µg/g | 101 |
| Zn | µg/g | 42 |
| Ga | µg/g | 5.7 |
| As | µg/g | 1.7 |
| Se | µg/g | 9.0 |
| Rb | µg/g | 2.8 |
| Sr | µg/g | 8.0 |
| Y | µg/g | 2.3 |
| Zr | µg/g | 7.6 |
| Nb | ng/g | 530 |
| Mo | µg/g | 2.1 |
| Pd | µg/g | 1.2 |
| Ag | ng/g | 53 |
| Sn | ng/g | 270 |
| Sb | ng/g | 87 |
| Te | ng/g | 440 |
| Cs | ng/g | 102 |
| Ba | µg/g | 2.7 |
| La | ng/g | 285 |
| Ce | ng/g | 763 |
| Pr | ng/g | 118 |
| Nd | ng/g | 547 |
| Sm | ng/g | 193 |
| Eu | ng/g | 70.7 |
| Gd | ng/g | 259 |
| Tb | ng/g | 52 |
| Dy | ng/g | 275 |
| Ho | ng/g | 68.9 |
| Er | ng/g | 208 |
| Tm | ng/g | 34.2 |
| Yb | ng/g | 200 |
| Lu | ng/g | 37.8 |
| Hf | ng/g | 170 |
| Re | ng/g | 90 |
| Ir | ng/g | 940 |
| Pt | µg/g | 1.8 |
| Th | ng/g | 39 |
| U | ng/g | 8.3 |

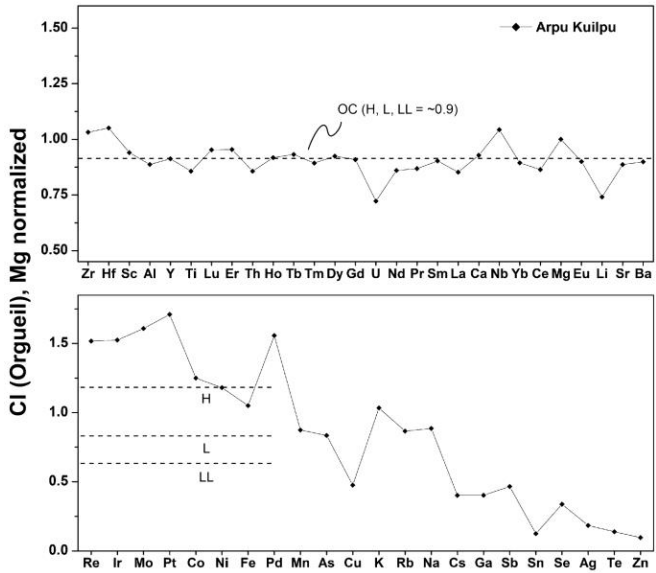

**Figure 4**. The bulk elemental concentrations of the Arpu Kuilpu meteorite, measured via ICP-MS, normalized to Magnesium and CI chondrite abundances. The top plot shows most of the lithophile elements, which have a mean and 1 σ normalized abundance of 0.91 ± 0.07, with the dotted line representing the 0.9 mean lithophile abundance observed in ordinary chondrites (Kallemeyn et al., 1989). The bottom plot displays Arpu Kuilpu's siderophile elemental concentrations (left, Re-Pd), as well as a few other lithophile and chalcophile abundances. The three dotted lines, represent the normalized mean siderophile abundances for H, L, and LL chondrites reported in Kallemeyn et al. (1989), which for Arpu Kuilpu is 1.42 ± 0.23, closest to that of H chondrites.

*Oxygen Isotopes*

The oxygen isotope ratios measured in Arpu Kuilpu can be seen in **Figure 5** and are as follows (all with 1σ uncertainties): $\delta^{17}O‰ = 2.904 ± 0.177$; $\delta^{18}O‰ = 4.163 ± 0.336$; $\Delta^{17}O‰ = 0.740 ± 0.002$.

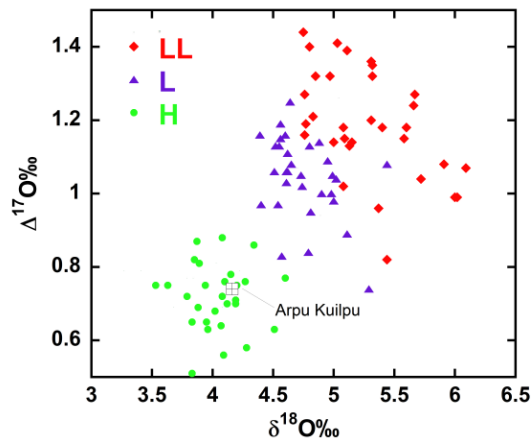

**Figure 5.** The oxygen isotope composition of Arpu Kuilpu plotted in the context of other ordinary chondrites (data sourced from Clayton et al. (1991)), clearly showing that this meteorite is an H chondrite.

### Laser Scanning and Ideal Gas Pycnometry

We measured the bulk density of the 14.64 g sample of Arpu Kuilpu to be $3.66 \pm 0.02$ g cm$^{-3}$, and a grain density of $3.77 \pm 0.02$ g cm$^{-3}$. These data taken together yield a porosity of $3.0\% \pm 0.7$, which is typical of ordinary chondrites (Britt and Consolmagno, 2003).

### Magnetic Susceptibility

The magnetic susceptibility measured from Arpu Kuilpu is log χ = $5.16 \pm 0.08$ (log SI units), consistent with other H chondrites (Consolmagno et al. 2006; Macke et al. 2011).

### Micro X-Ray Computed Tomography
#### Bulk Density

Using the μCT data from the 430.3 mg piece, we measured the subsample's volume to be 118.5 mm$^3$, yielding a bulk density of 3.63 g cm$^{-3}$ which is in good agreement with the 3.66 g cm$^{-3}$ found by laser scanning and Ideal Gas Pycnometry of a separate aliquot (see above). This indicates that the sample is relatively homogeneous at the 0.1-10 cm$^3$ scale. Visual inspection of the tomography volumes confirms this observation. These bulk densities are consistent with the average H chondrite fall bulk density of $3.42 \pm 0.19$ g cm$^{-3}$ (Consolmagno et al. 2008).

#### Petrofabric and Shock State

A typical μCT "slice" of the Arpu Kuilpu ordinary chondrite is shown in **Figure 6**. The petrofabric, as defined by the metal grain shape preferred orientations (see methods), of Arpu Kuilpu shows the major axes of the grains define a relatively indistinct girdle and the minor axes cluster weakly at high angles to the major axis girdle (**Figure 7)**. Together these features are the signature of a foliation petrofabric. It is generally agreed that foliation petrofabrics in chondrites are the result of grain rotation and alignment during uniaxial hypervelocity impact deformation (Gattacceca et al. 2005; Friedrich et al. 2008).

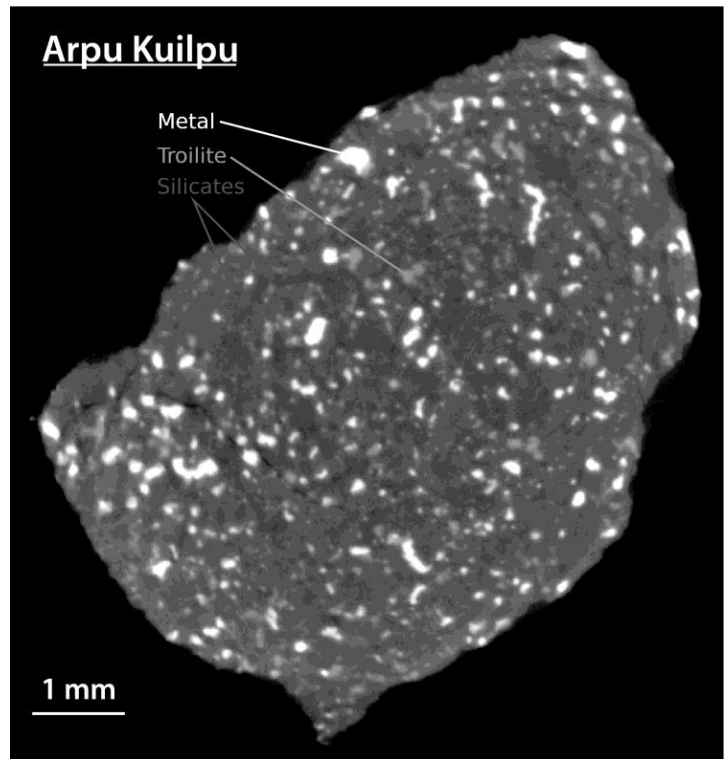

**Figure 6**. μCT "slice" of a chip of the Arpu Kuilpu ordinary chondrite. The higher the greyscale value, the higher the average atomic weight of the material. The bright white Fe-Ni metal grains and pale grey Fe sulfides can be easily distinguished from the moderate grey silicates and the air around the chip represented as black. Although small cracks can occasionally be seen, there is no evidence of brecciation.

Examination of the eigenvalue tensor fabric descriptors shows that the major axis shape parameter ($K_{major}$ = 0.73) indicates a girdle distribution (K<1) (also see **Figure 7**). The minor axis shape parameter ($K_{minor}$ = 1.401) can be classified as a cluster distribution (K>1). Both of these parameters agree with visual inspection of the fabric. The major axis anisotropy parameter has a value of 0.071. It is known that the fabric anisotropy varies in intensity with increasing petrographic shock stage (Friedrich et al. 2008; Friedrich et al. 2017). Based on an anisotropy of 0.071, a best estimate for the shock stage (Stöffler et al. 2018) of the sample may be S1 to S2, which matches with other microstructural features observed in the thin section. We also examined the anisotropy of the fabric of the largest 1000 grains in the smaller 435 mg sample collected with a different CT instrument and at a different spatial resolution. The anisotropy parameter for this sample was strikingly similar with a value of 0.073.

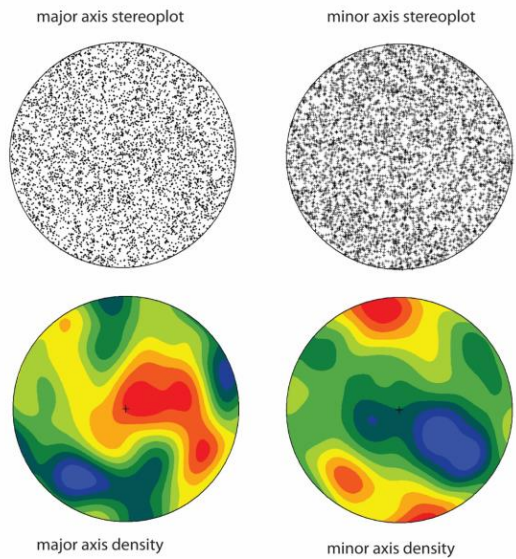

**Figure 7**. Stereoplots and densities of orientations of the major and minor axis of best fit ellipsoids around the largest 5000 metal grains in the Arpu Kuilpu ordinary chondrite. Arpu Kuilpu's major axis fabric is a relatively indistinct girdle and the minor axis is a cluster, together being the signature of a very weak foliation petrofabric.

The shapes of the digitally constructed best-fit ellipsoids digitally constructed around the largest 5000 metal grains in Arpu Kuilpu are shown in **Table 5**. Of the metallic grains, 35.7 % have sub-equant spheroid shapes (see Blott and Pye, 2008 for shape descriptors) with the next most abundant (31.6 %) being prolate spheroids. Oblate spheroid grains (16.4%) make up the largest remaining category of metal grain shapes.

**Table 5.** Shapes of largest 5000 metal grains in the Arpu Kuilpu chondrite

| Shape | n | % |
|---|---|---|
| Equant Spheroid | 97 | 1.9 |
| Sub-Equant Spheroid | 1783 | 35.7 |
| Prolate Spheroid | 1579 | 31.6 |
| Roller | 331 | 6.6 |
| Blade | 340 | 6.8 |
| Oblate Spheroid | 819 | 16.4 |
| Very Oblate Spheroid | 51 | 1.0 |
| Discoid | 0 | 0.0 |

In **Figure 8**, we place the physical properties porosity and fabric anisotropy of the Arpu Kuilpu into the context of other ordinary chondrites. Arpu Kuilpu falls into a region of very low porosity, but also very low fabric anisotropy that is indicative of high ambient heat during or after the minor shock pressure that Arpu Kuilpu experienced (see Friedrich et al. 2017).

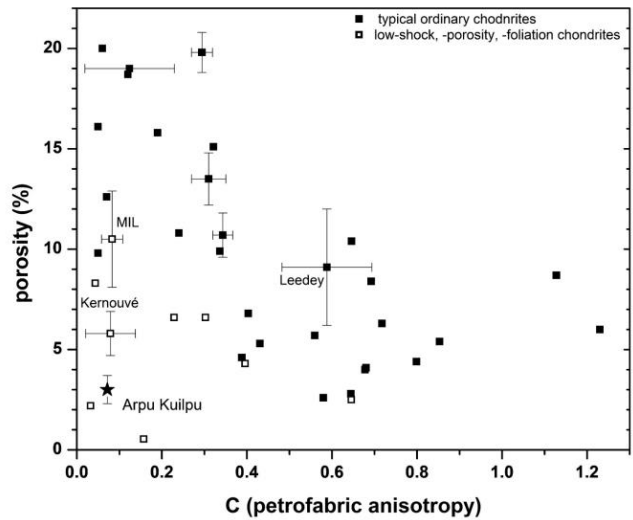

**Figure 8.** Plot of porosity versus petrofabric anisotropy for typical ordinary chondrites and a suite of low-shock, low-porosity, low-foliation ordinary chondrites (data from Friedrich et al. 2017) and the Arpu Kuilpu ordinary chondrite. Errors of these measured values are shown for selected samples to give the reader an idea of the typical magnitude of the associated errors. Arpu Kuilpu falls into a region of very low porosity, but also very low fabric anisotropy, indicative of the sample having experienced high ambient heat during or after the impact-related shock that has affected this meteorite.

*Gamma-Ray Spectrometry*

The results from the gamma-ray spectrometry measurements displayed in **Table 6** were briefly mentioned by Shober et al. (2022) but are re-presented here for clarity. The main conclusion of Shober et al. (2022) was that the low $^{60}$Co activity (0.8 ± 0.3 dpm kg$^{-1}$) indicated that the original meteoroid was small enough (~10 cm; Bonino et al. 2001) that its recent cosmic-ray exposure in space did not produce a significant flux of thermal neutrons. The measured $^{26}$Al activity using this method is consistent with a small H chondrite (Bhandari et al. 1989; Bonino et al. 2001; Leya and Masarik, 2009), and with the AMS measurements (see below) of $^{26}$Al. Considering the time of measurement (183 days after the fireball event) the activities are consistent with a rock that stopped accumulating cosmogenic radionuclides on 1 June 2019.

**Table 6.** Massic activities (corrected to date of fall of the meteorite 1st June 2019) of cosmogenic and primordial radionuclides in the specimen of the Arpu Kuilpu stone measured by non-destructive gamma-ray spectrometry. Errors include a 1σ uncertainty of 10% in the detector efficiency calibration.

| Cosmogenic Radionuclides | Half-life | Arpu Kuilpu (21.8 g) |
|---|---|---|
| $^7$Be | 53.22 d | 110 ± 30 dpm kg$^{-1}$ |
| $^{58}$Co | 70.83 d | 4 ± 2 dpm kg$^{-1}$ |
| $^{56}$Co | 77.236 d | 6 ± 2 dpm kg$^{-1}$ |
| $^{46}$Sc | 83.787 d | 10 ± 2 dpm kg$^{-1}$ |
| $^{57}$Co | 271.8 d | 8 ± 1 dpm kg$^{-1}$ |
| $^{54}$Mn | 312.3 d | 51 ± 4 dpm kg$^{-1}$ |
| $^{22}$Na | 2.60 y | 62 ± 5 dpm kg$^{-1}$ |
| $^{60}$Co | 5.27 y | 0.8 ± 0.3 dpm kg$^{-1}$ |
| $^{44}$Ti | 60 y | < 4.3 dpm kg$^{-1}$ |
| $^{26}$Al | 7.17x10$^5$ y | 32 ± 3 dpm kg$^{-1}$ |
| Nuclide | | |
| U | | 10 ± 1 ng g$^{-1}$ |
| Th | | 50 ± 4 ng g$^{-1}$ |
| K | | 700 ± 70 ng g$^{-1}$ |

When we compare the radionuclide concentrations with cosmic ray production estimations for $^{26}$Al (Leya and Masarik, 2009), $^{60}$Co (Eberhardt et al. 1963, Spergel et al. 1986), $^{54}$Mn (Kohman and Bender, 1967), and $^{22}$Na (Bhandari et al. 1993), and assume the specimen is from the central part (which may not be a reliable assumption, see Discussion), the best agreements are obtained (in the sequence of the given isotopes) for radii of < 10 cm, < 20 cm, < 5 cm and 5-8 cm. The $^{22}$Na/$^{26}$Al ratio for this specimen is 1.9 ± 0.2. Combining these results of these radionuclides, we infer a radius of 5 to 10 cm, for a roughly spherical meteoroid. Alternatively, it can be from the surface (<5 cm) of a larger meteoroid.

The concentrations of U, Th, and K, derived from the activities of the naturally occurring nuclides ($^{235}$U, $^{238}$U, $^{232}$Th, $^{40}$K; **Table 6)** are consistent with average concentrations in H chondrites (Wasson and Kallemeyn, 1988).

### *Long-lived Cosmogenic Radionuclides (AMS and ICP-OES)*

In general, the bulk composition of the Arpu Kuilpu sample that was used for cosmogenic radionuclide analysis (measured via ICP-OES, see **Table 7**) is consistent with both the average H chondrite composition of Wasson and Kallemeyn (1988), and with the measurements made via ICP-MS (see above). The Co and Ni concentrations in the metal fraction of Arpu Kuilpu are consistent with H chondrite classification, while the Mg concentration of <10 ppm indicates that the purified metal contained no detectable (<0.01 wt%) silicate contamination. This implies that $^{10}$Be and $^{26}$Al contributions from silicates are negligible and thus that the measured $^{10}$Be and $^{26}$Al concentrations require no correction.

**Table 7.** Chemical composition of metal and stone fractions of Arpu Kuilpu. All values are in wt %. The bulk composition (ICP-OES) is calculated from measured values in metal and stone fractions, which are 13.9 and 86.1 wt%, respectively. The ICP-MS measurements (see **Table 4)** are repeated here for comparison. The last column shows average H chondrite composition from Wasson and Kallemeyn (1988). Entries 'nm' indicate not measured.

| Element | Arpu Kuilpu (H) | | | | H chondrite |
|---|---|---|---|---|---|
| | Metal | Stone | Bulk (ICP-OES) | Bulk (ICP-MS) | |
| Mg | <0.001 | 15.4 | 13.2 | 13.5 | 14.0 |
| Al | nm | 1.26 | 1.08 | 1.09 | 1.13 |
| P | nm | 0.11 | 0.10 | nm | 0.11 |
| S | nm | 1.91 | 1.64 | nm | 2.0 |
| K | nm | 0.12 | 0.10 | 0.083 | 0.08 |
| Ca | nm | 1.53 | 1.32 | 1.19 | 1.25 |
| Ti | nm | 0.071 | 0.061 | 0.071 | 0.060 |
| Mn | nm | 0.265 | 0.228 | 0.23 | 0.232 |
| Fe | 89.9 | 14.6 | 25.1 | 27.6 | 27.5 |
| Co | 0.470 | 0.013 | 0.076 | 0.087 | 0.081 |
| Ni | 8.5 | 0.52 | 1.63 | 1.84 | 1.60 |

The results of the AMS measurements of Arpu Kuilpu are shown in **Table 8.** The high concentrations of $^{10}$Be and $^{26}$Al in the metal phase as well as the $^{26}$Al/$^{10}$Be ratio of 0.77 ± 0.03 indicate that the Arpu Kuilpu meteorite had a minimum CRE age of 4-6 Myr in a relatively small object (radius < 10 cm).

**Table 8**. Cosmogenic radionuclide concentrations (in dpm/kg) in the metal and stone fractions of Arpu Kuilpu. The bulk values are calculated from the metal and stone based on measured metal and stone proportions of 13.9 and 86.1 wt%, respectively. The last column shows the $^{10}$Be(sto)/$^{10}$Be(met) and $^{26}$Al(sto)/$^{26}$Al(met) ratios, which can be used as a shielding indicator (see Discussion).

| Nuclide | Metal | Stone | Bulk | Sto/Metal |
|---|---|---|---|---|
| $^{10}$Be | 5.58 ± 0.16 | 16.10 ± 0.25 | 14.6 ± 0.2 | 2.89 ± 0.09 |
| $^{26}$Al | 4.30 ± 0.12 | 42.6 ± 0.8 | 37.3 ± 0.7 | 9.9 ± 0.3 |
| $^{36}$Cl | 19.8 ± 0.5 | 5.02 ± 0.13 | 7.07 ± 0.1 | N/A |

*Noble Gas Isotopes*

The measured concentrations and ratios of the noble gases in Arpu Kuilpu are listed in **Table 9.** Both He and Ne are purely cosmogenic (cos) apart from $^4$He. The non-cosmogenic fraction of the $^4$He signal is assumed to be radiogenic (rad), which is supported by the Ne isotope values that do not indicate the presence of a trapped (tr) component. A solar wind component could therefore be excluded, which also rules out the possibility of this meteorite being a regolith breccia. The absence of primordially trapped He is likely caused by thermal processing at temperatures going up to ~740 °C for H5 chondrites (olivine–spinel equilibrium temperature, Kessel et al. 2007).

The $^{36}$Ar/$^{38}$Ar values for both samples (3.28 ± 0.02 for AKS and 2.73 ± 0.02 for AKL) do not match pure cosmogenic compositions (0.63-0.67; Wieler, 2002) and imply mixing between trapped (tr) and cosmogenic (cos) Ar. The trapped component cannot be identified since the isotopic compositions Q and air $^{36}$Ar/$^{38}$Ar ratios are too close to each other. Although a small cosmogenic component can be observed in $^{38}$Ar, it can neither be recognized in the $^{78}$Kr/$^{84}$Kr and $^{80}$Kr/$^{84}$Kr ratios nor in the light Xe isotopes. Both measured aliquots display high $^{129}$Xe/$^{132}$Xe ratios resulting from a short-lived $^{129}$I-derived $^{129}$Xe excess, a feature commonly observed in OCs of types 5 or 6 (e.g. Alaerts et al. 1979; Moniot, 1980), fitting with Arpu Kuilpu's H5 petrologic type.

**Table 9.** The noble gas isotopic concentrations and ratios of the two measured aliquots from Arpu Kuilpu (AKL and AKS; $^{84}$Kr = 100; $^{132}$Xe = 100).

|  | AKS | AKL |
|---|---|---|
| **Mass [mg]** | 27.178 ± 0.014 | 29.267 ± 0.015 |
| $^4$He 10$^{-8}$ cm$^3$ STP/g | 1752.9 ± 5.2 | 1625.4 ± 6.8 |
| $^3$He/$^4$He 10$^{-4}$ | 49.4 ± 1.2 | 53.37 ± 0.71 |
| $^{20}$Ne 10$^{-8}$ cm$^3$ STP/g | 0.9845 ± 0.0090 | 1.047 ± 0.010 |
| $^{20}$Ne/$^{22}$Ne | 0.8107 ± 0.0044 | 0.8144 ± 0.0046 |
| $^{21}$Ne/$^{22}$Ne | 0.7776 ± 0.0030 | 0.7658 ± 0.0050 |
| $^{36}$Ar 10$^{-8}$ cm$^3$ STP/g | 1.4919 ± 0.0053 | 0.9862 ± 0.0056 |
| $^{36}$Ar/$^{38}$Ar | 3.280 ± 0.018 | 2.728 ± 0.017 |
| $^{40}$Ar/$^{36}$Ar | 3738 ± 18 | 6148 ± 41 |
| $^{84}$Kr 10$^{-10}$ cm$^3$ STP/g | 1.299 ± 0.014 | 1.013 ± 0.020 |
| $^{78}$Kr/$^{84}$Kr | 0.585 ± 0.086 | 0.31 ± 0.26 |
| $^{80}$Kr/$^{84}$Kr | 4.11 ± 0.15 | 3.95 ± 0.25 |
| $^{82}$Kr/$^{84}$Kr | 20.37 ± 0.40 | 21.25 ± 0.54 |
| $^{83}$Kr/$^{84}$Kr | 21.57 ± 0.33 | 22.09 ± 0.56 |
| $^{86}$Kr/$^{84}$Kr | 31.15 ± 0.67 | 31.87 ± 0.85 |
| $^{132}$Xe 10$^{-10}$ cm$^3$ STP/g | 2.129 ± 0.028 | 1.451 ± 0.020 |
| $^{124}$Xe/$^{132}$Xe | 0.490 ± 0.021 | 0.442 ± 0.023 |
| $^{126}$Xe/$^{132}$Xe | 0.400 ± 0.019 | 0.441 ± 0.036 |
| $^{128}$Xe/$^{132}$Xe | 8.32 ± 0.11 | 8.33 ± 0.14 |
| $^{129}$Xe/$^{132}$Xe | 123.33 ± 0.60 | 140.52 ± 0.87 |
| $^{130}$Xe/$^{132}$Xe | 15.95 ± 0.13 | 16.44 ± 0.15 |
| $^{131}$Xe/$^{132}$Xe | 80.87 ± 0.54 | 81.70 ± 0.63 |
| $^{134}$Xe/$^{132}$Xe | 37.98 ± 0.33 | 38.75 ± 0.46 |
| $^{136}$Xe/$^{132}$Xe | 31.79 ± 0.31 | 32.19 ± 0.29 |

The U/Th-He and K-Ar chronometers for both aliquots yield old gas retention ages around 4.0-4.6 Gyr essentially within the same range (**Table 10**), suggesting no major resetting events in Arpu Kuilpu's history.

**Table 10.** Radiogenic concentrations of He and Ar (in 10$^{-8}$ cm$^3$ STP/g), and gas retention ages (T$_4$ and T$_{40}$) calculated using U, K, and Th concentrations from the Gamma-ray spectroscopy section of this study (see **Table 6**).

|  | **AKS** | **AKL** |
|---|---|---|
| $^4$He$_{rad}$ | 1704.0 ± 6.6 | 1576.4 ± 7.9 |
| $^{40}$Ar$_{rad}$ | 5380 ± 200 | 5940 ± 140 |
| T$_4$ (Gyr) | 4.2-4.6 | 4.0-4.4 |
| T$_{40}$ (Gyr) | 4.4-4.8 | 4.5-4.7 |

The production rates of cosmogenic $^3$He, $^{21}$Ne and $^{38}$Ar calculated with the above-mentioned model and the resulting CRE ages of the two aliquots are given in **Table 11**. The T$_3$, T$_{21}$ and T$_{38}$ ages are comparable, with the T$_{21}$ and T$_{38}$ ages overlapping within their ranges for AKS. The T$_3$ ages for both aliquots overlap with each other and are lower than the T$_{21}$ and T$_{38}$ ages. The T$_{21}$ age for AKL is slightly higher than for AKS. This discrepancy could stem from a difference in exposure to solar cosmic rays (SCR) which can affect material in the upper few cm of a meteoroid. While the $^{22}$Ne/$^{21}$Ne ratios of both aliquots suggest a pre-atmospheric radius of the meteoroid of 1 cm according to the model, the sample depth of AKL is shallower, which might have resulted in a slightly higher SCR-derived $^{21}$Ne$_{cos}$ production (**Table 11**). Due to this possibility, the T$_{21}$ age for AKL of ~10 Myr is not used as the maximum for our preferred CRE age. The ($^{3}$He/$^{21}$Ne)$_{cos}$ and ($^{21}$Ne/$^{22}$Ne)$_{cos}$ ratios of AKS and AKL do not indicate significant $^3$He loss. According to the cosmogenic noble gas concentrations, the CRE age for Arpu Kuilpu is ~7-9 Myr (**Table 11**).

**Table 11.** Suitable shielding conditions matching our cosmogenic $^{21}$Ne/$^{22}$Ne value (average for AKS and AKL: 0.772 ± 0.006), determined with the model by Leya and Masarik (2009) updated for small H chondrite meteoroids (Wieler et al. 2016), cosmogenic $^3$He, $^{21}$Ne and $^{38}$Ar isotope concentrations (in $10^{-8}$ cm$^3$ STP/g), production rates P$_x$ (in $10^{-8}$ cm$^3$/(g × Myr)), and calculated CRE ages T$_x$ (in Myr).

| *Sample* | AKS | AKL |
|---|---|---|
| **Meteoroid Radius (cm)** | 1 | 1 |
| **Burial Depth (cm)** | 0.35-0.95 | 0.00-0.15 |
| $^3$He$_{cos}$ | 8.66 ± 0.22 | 8.67 ± 0.11 |
| $^{21}$Ne$_{cos}$ | 0.9444 ± 0.0078 | 0.9849 ± 0.0100 |
| $^{38}$Ar$_{cos}$ | 0.1993 ± 0.0030 | 0.2010 ± 0.0030 |
| P$_3$ | 1.20-1.26 | 1.22-1.23 |
| P$_{21}$ | 0.106-0.113 | 0.10-0.11 |
| P$_{38}$ | 0.0237-0.0248 | 0.0239-0.02340 |
| T$_3$ | ~7 | ~7 |
| T$_{21}$ | ~8-9 | ~9-10 |
| T$_{38}$ | ~8-9 | ~8 |
| **Preferred CRE Age (Myr)** | ~7-9 | ~7-9 |

### *$^{40}$Ar/$^{39}$Ar Chronology*

We selected nine pyroxene aliquots including between 1 and 30 grains, and with each grain ranging in size from 150-350 µm in diameter, along with one plagioclase grain for $^{40}$Ar/$^{39}$Ar chronology. Seven pyroxene aliquots yielded statistically indistinguishable plateau ages ranging from 4409 ± 71 Ma and 4501 ± 178 Ma (**Figure 9B**) and their combined ages result in a weighted mean age of 4467 ± 16 Ma (**Figure 9A**; P = 0.16; 2σ). Two single grain pyroxene analysis yielded plateau age of 4251 ± 163 Ma and 4310 ± 135 Ma slightly younger than the pyroxene bulk population. The plagioclase analysis did not yield enough gas for a successful analysis.

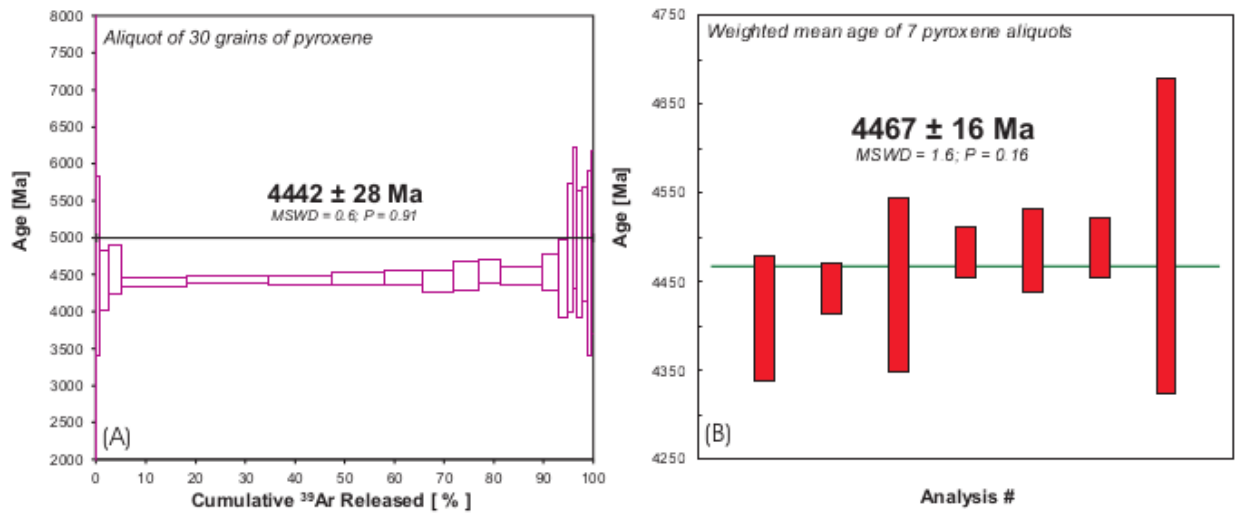

**Figure 9.** (A) Example of a plateau age (2σ) obtained for one (Hughes 30GRN 150u pyxB) of the seven pyroxene aliquots extracted from the Arpu Kuilpu meteorite. (B) Weighted mean age of the seven plateau ages obtained on pyroxene populations.

## DISCUSSION

### *Meteorite Class and Type*
Every relevant analysis we have performed on this meteorite confirms that Arpu Kuilpu is indeed an H5 ordinary chondrite. Gamma-Ray spectrometry, µCT, oxygen isotopes, ICP-MS, ICP-OES, Mössbauer spectroscopy, ideal gas pycnometry, and magnetic susceptibility, each indicate the general H chondrite meteorite class, while results from the SEM-EDS (**Figure 1)** and optical microscopy (**Figure 2**) analyses reveal a texture consistent with a petrologic type 5, further confirmed by the EPMA measurements of individual mineral compositions. The lack of trapped noble gasses also excludes the possibility for lower (3 or 4) petrologic types.

### *Weathering*
Arpu Kuilpu was recovered approximately six weeks after its fall, and although 8.4 mm of rainfall was recorded at the nearest weather station at Forrest Airport in Western Australia between its fall and recovery (Shober et al. 2022), visual inspection of the meteorite did not reveal significant weathering. Inspection of the thin section via optical microscopy also reveals little terrestrial alteration, all of which is also supported by a very minor $Fe^{3+}$ signal from the Mössbauer spectroscopy measurements (**Table 3; Figure 3**).

### *Shock State and Impact History*
The traditional method of determining the shock stage in meteorites, optical microscopy, has revealed that Arpu Kuilpu experienced relatively little shock (S1-S2; Stöffler et al. 2018). This is further supported by petrofabric measurements taken from the µCT data (Friedrich et al. 2008; 2017), seen in **Figure 7**. By combining the major axis anisotropy value of C = 0.071 with the porosity of 3.0 ± 0.7% (measured via ideal gas pycnometry), as seen in **Figure 8**, Arpu Kuilpu likely experienced high ambient heat at approximately the same time that it underwent its minor impact induced shock. The results of the $^{40}Ar/^{39}Ar$ experiments precisely constrain the latest possible time for such an impact event to the very early solar system at 4467 ± 16 Ma, which is also supported by old ages (> 4.0 Gyr) from the noble gas analysis via U/Th-He dating, further ruling out the possibility for recent impacts (see **Table 10**).

Note that the $^{40}Ar/^{39}Ar$ systematic in pyroxene is relatively easy to reset and even minor impacts can produce enough energy to reset pyroxene provided that the shock wave creates a spike of high temperature for duration as short as few micro-seconds (Cassata et al., 2011; Kennedy et al., 2019; Jourdan et al., 2020), which in the present case, would be exacerbated by a high ambient heat. Minor subsequent impacts might have been recorded by individual crystals of pyroxene at 4251 ± 163 Ma and 4310 ± 135 Ma, but considering that these two ages are isolated, the shock event might have been very small with very focused energy, not sufficient to affect the majority of the pyroxene crystals.

### *Meteoroid Size, Burial Depth, and CRE Age*

Determining the irradiation depth of Arpu Kuilpu within its original meteoroid, as well as the meteoroid's size, and CRE age, requires the insight of three methodologies employed in this study: gamma-ray spectrometry, AMS, and noble gas analysis, as well as the fireball observations presented in Shober et al. (2022). For the pre-atmospheric size of the meteorite, each method provides an estimate based on a different time interval, with the fireball data representing the size at the time of atmospheric entry, the short-lived radionuclides the size during the past few years before entry, the long-lived radionuclide the size of the past few million years before entry and the noble gases the average size during its entire cosmic-ray exposure. For a meteorite with a simple CRE history, i.e., with no significant changes in size during CRE, all four estimates should yield the same size within the uncertainties of each method and the model calculations associated with it.

The first calculation of Arpu Kuilpu's original size as a meteoroid is given in Shober et al. (2022), where a pre-atmospheric size of approximately 5 cm (radius) is derived from the fireball observations, though this size estimate is prone to variation due to uncertainties in model parameters such as shape and spin of the meteoroid which are assumed values. Arpu Kuilpu's atmospheric entry was also observed to have experienced a fragmentation event near the end of the fireball (Shober et al. 2022), further opening the possibility that the meteorite could have come from a non-central part of the meteoroid.

The results of the non-destructive HPGe Gamma-Ray Spectrometry measurements (**Table 6**) also support a small meteoroid size between 5-10 cm (assuming a spherical meteoroid). The activities of the short-lived radioisotopes, with half-lives that are less than the orbital period (**Table 6**), represent the production integrated over the last segment of the orbit. The fall of Arpu Kuilpu occurred during a minimum at the end of solar cycle 24, as indicated by the neutron monitor data (Bartol, 2020). The cosmic ray flux was high in the six months before the fall, so the activities for the very short-lived radionuclides are expected to be high, as earlier reported (Jenniskens et al. 2014 and references cited therein; see **Table 6**).

As discussed in the methods, the $^{22}Ne/^{21}Ne$ ratio can be used as a shielding factor to estimate the size of the original meteoroid, the cosmogenic noble gas production rates, and the meteorite's depth within the meteoroid. For Arpu Kuilpu's two analyzed aliquots, the $^{22}Ne/^{21}Ne$ ratios nearly agree within error (~1.28; see **Table 9**), and are relatively high, suggesting a solar cosmic ray (SCR) component. As stated in the Methods and Results sections, these high $^{22}Ne/^{21}Ne$ ratios yielded no matches for meteoroids 10-500 cm when using the model from Leya and Masarik (2009), so we instead used a modified version of this model (Wieler et al. 2016) intended for a meteoroid radius up to 7 cm, which yields a very small meteoroid radius (~1 cm), and an even shallower burial depth (<1 cm) for Arpu Kuilpu. Considering the modeled production rates of $^{3}He$, $^{21}Ne$, and $^{38}Ar$ (**Table 11**), the noble gas content suggests a CRE age of ~7-9 Ma.

The AMS-measured concentrations of $^{10}Be$ (16.1 ± 0.2 dpm kg$^{-1}$) and $^{26}Al$ (42.6 ± 0.8 dpm kg$^{-1}$) (see **Table 7)** in the stone fraction of Arpu Kuilpu are on the low end of the range of 16-25 dpm kg$^{-1}$ ($^{10}$Be) and 40-85 dpm kg$^{-1}$ ($^{26}$Al) predicted by the model of Leya and Masarik (2009), suggesting that Arpu Kuilpu's meteoroid was small with maximum radius of 10 cm (**Figure 10**). Unfortunately, this model did not provide production rates for objects smaller than 10 cm, though these concentrations of $^{10}$Be, $^{26}$Al and $^{36}$Cl are consistent with irradiation at a depth of <1 cm in a 10 cm object (**Figure 10** 1b-1d). The very low shielding conditions are also supported by the relatively high $^{10}$Be and $^{26}$Al concentrations in the metal phase and the low $^{10}$Be(sto)/$^{10}$Be(met) and $^{26}$Al(sto)/$^{26}$Al(met) ratios of 2.9 and 9.9, respectively (**Table 7**). These ratios are strongly dependent on shielding conditions (Nagai et al. 1990), with $^{10}$Be(sto)/$^{10}$Be(met) increasing from values of ~3 in small chondrites to values of 5-13 in large chondrites, like FRO 90174 and Gold Basin (Welten et al. 2001; 2003). The $^{26}$Al(sto)/$^{26}$Al(met) ratios show a range of 9-70 for the same chondrites. The values found in Arpu Kuilpu are among the lowest values reported in ordinary chondrites, similar to those found in several small Antarctic H-chondrites from the Frontier Mountain (FRO) icefield (Welten et al. 2001), including FRO 90037 (13.1 g) and 90151 (33.6 g). These small H-chondrites also had very high $^{22}$Ne/$^{21}$Ne ratios of 1.27-1.32, similar to Arpu Kuilpu (1.28, see **Table 9**). Such little shielding appears unlikely, as the outer >1 cm of the meteoroid is often lost during atmospheric ablation. However, since fragmentation was observed during the fireball event, it is quite possible that the meteorite came from a location close to the surface of the meteoroid, and survived to the ground, experiencing non-uniform ablation. Wieler et al. (2016) and Roth et al. (2017) have shown that SCR-derived Ne can commonly be observed in small ordinary chondrites and martian meteorites. An alternative approach to calculating the CRE age, which does not rely on the $^{22}$Ne/$^{21}$Ne ratio, is to use the correlation between the $^{3}$He, $^{21}$Ne, and $^{38}$Ar production rates as well as the bulk $^{26}$Al production rate from the Leya and Masarik (2009) model albeit only for meteoroids with radii ≥ 10 cm. Arpu Kuilpu's bulk $^{26}$Al concentration of 37.2 dpm kg$^{-1}$ is just within the 35-70 dpm kg$^{-1}$ range found in this model. Assuming that these correlations can be extended to slightly smaller objects with ~5 cm radius, we can estimate $^{3}$He, $^{21}$Ne and $^{38}$Ar production rates (in units of 10$^{-8}$ cm$^{3}$ STP/g/Ma) of 1.40, 0.176 and 0.035, respectively (note that these differ from those presented in **Table 11,** which used the modified model from Leya and Masarik (2009) extended to radii ≤10 cm (Wieler et al. 2016) and $^{22}$Ne/$^{21}$Ne ratios). Using the cosmogenic $^{3}$He, $^{21}$Ne and $^{38}$Ar concentrations from **Table 11**, this approach yields CRE ages of 6.2, 5.5 and 5.7 Ma, with an average value of 5.8 ± 0.6 Ma (based on an uncertainty of ~10% for the production rate calculations).

This slight discrepancy in results between the noble gas only-informed model for small radii ≤10 cm (Wieler et al. 2016; 1 cm radius, <1 cm burial depth, ~7-9 Ma CRE age), and the $^{26}$Al-informed Leya and Masarik (2009) approach using a model for radii ≥ 10 cm, extending it to ≤10 cm (adopting a 5-10 cm radius (based on gamma-ray spectrometry and short-lived radionuclides), <1 cm burial depth, ~6 Ma CRE age), could be explained in two ways: sample heterogeneity and model uncertainty. Although the aliquots for the AMS, noble gas, and ICP-MS analyses all originated from the same physical location, approximately ~0.5 cm within the meteorite, they sometimes show minor, yet noticeable deviation from each other and from typical H chondrite elemental concentrations. The ICP-MS results, for instance, show an enrichment in trace siderophile elements which we attribute to an embedded refractory siderophile nugget (**Figure 4; Table 4**). While the noble gas concentrations for the two aliquots measured (AKS and AKL) are close to one another, they often do not agree within error (**Table 9**). Considering these aliquot-to-aliquot variations, and model uncertainties in both Wieler et al. (2016), and Leya and Masarik (2009), particularly concerning the covered meteoroid size, we assert these differences to be at the edge of our ability to infer the exact meteoroid history, and that future work may be needed to supplant or expand on existing models to comprehensively extend our understanding of CRE to very small (<10 cm) meteoroids and their surviving meteorites.

Combining these methods and their insights, we conclude that Arpu Kuilpu came from a small (<5 cm) meteoroid, very near to its surface (<1 cm), and that its meteoroid spent ~6-8 Ma in interplanetary space, separate from its contemporary asteroidal source body.

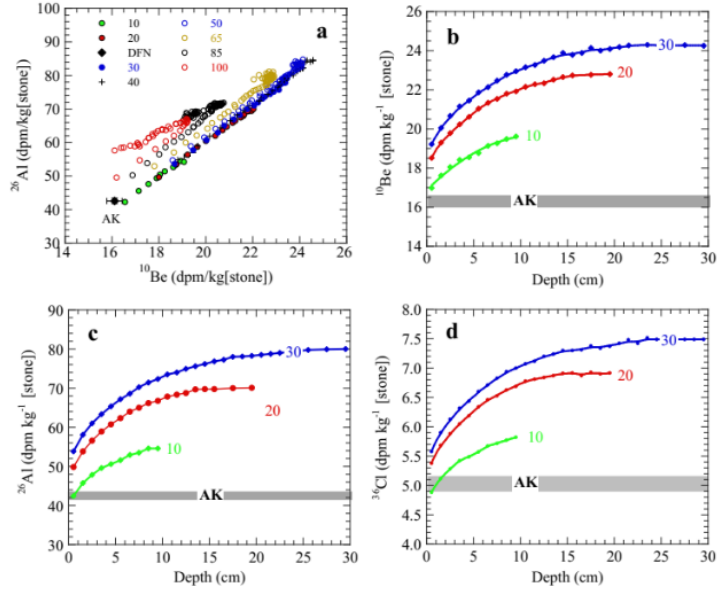

**Figure 10.** (a) Comparison of the measured concentrations of $^{10}$Be and $^{26}$Al in the stone fraction of Arpu Kuilpu with calculated $^{10}$Be and $^{26}$Al production rates in the stone fraction of ordinary chondrites with radii of 10-100 cm. Panels (b-d) shows a comparison of the measured concentrations of $^{10}$Be (b) $^{26}$Al (c) and $^{36}$Cl (d) in the stone fraction of Arpu Kuilpu with calculated depth profiles in the stone fraction of H-chondrites with radii of 10-30 cm. All calculations are based on the model of Leya and Masarik (2009).

### *Relation to Other Orbital Meteorites*

Considering the 53 other meteorites with associated orbits that have been recovered so far (at the time of this writing; Meier, 2017), the equilibrated H chondrites (H4s, H5s, and H6s) account for more than 30% (excluding multi-lithological falls), to which we will compare Arpu Kuilpu based on two main criteria: meteoroid orbit and CRE age (**Table 12**). We would also like to note that although the H5/6 chondrites Al-Khadhaf (MB 112; Gattacceca et al., 2024) and Santa Filomena (Tosi et al., 2023) are both falls with known orbits, their CRE ages and probable source resonances are currently unknown, therefore we did not include them in our comparisons to Arpu Kuilpu. For likeness in orbital parameters, Ejby (Spurný et al. 2017) and Košice (Borovička et al. 2013) are the closest matches as they both have a high probability of being sourced from a JFC orbit (although with relatively high associated uncertainties), with Hamburg (Brown et al. 2019), then Murrili (Sansom et al. 2020) being next closest with Tisserand parameters near the border of being considered a JFC orbit. Interestingly, the CRE ages of Murrili (7 ± 1 Myr; Anderson et al. 2021) and Košice (5-7 Myr; Povinec et al. 2015) are similar to Arpu Kuilpu's ~6-8 Myr, falling within the broad peak of 6-8 Myr for H5 chondrites identified by Marti and Graf (1992). Although the Lost City and Morávka meteorites also have CRE ages falling into this broad peak (Bogard et al. 1971; Borovička et al. 2003b), their orbits are distinctly dissimilar, originating from the inner main belt, on the other side of the 3:1 mean motion resonance with Jupiter (McCrosky et al. 1971; Borovička et al. 2003a). Therefore, at first assessment, the most likely orbital meteoritic sibling to Arpu Kuilpu is Košice, since it has the same CRE age and orbit, within errors (which are comparatively high for the latter).

Though to properly compare Arpu Kuilpu to other orbital H chondrites, we must also consider how orbits can evolve over time. Although the semi major axes of most small meteoroids change gradually over many orbital periods, due to the Yarkovsky effect (Bottke et al., 2001), eventually wandering into a mean motion resonance with Jupiter which then increases its eccentricity ultimately causing a collision with one of the inner planets, the work by Shober et al. (2020a; 2020b) demonstrates an additional mechanism for orbital evolution. They show how a meteoroid with a near-Earth orbit (a < 2 AU) can have a close encounter with an inner planet, immediately altering its path to a JFC-like orbit. This is further supported by a later study (Shober et al. 2021) that shows how most objects in the cm-m size range on JFC orbits are asteroidal in nature, and not composed of volatile-rich material (originating from trans-Neptunian space (Nesvorný et al. 2010; Fernández and Sosa, 2015)) typical of dust and km sized objects existing on similar orbits. Using this mechanism, we can speculate that Arpu Kuilpu, Košice, Murrili, and Lost City were all liberated from the same source body in the same ejection event (~6-8 Ma) in the inner solar system, but that only Arpu Kuilpu and Košice underwent close encounters to transfer into a JFC-like orbit. Although Morávka has a similar CRE age to Arpu Kuilpu, its high inclination (i = 32; **Table 12**), especially compared to the other H5's listed here, make a close orbital link unlikely. For Ejby and Hamburg, it is possible that they were ejected at different times, from either the same or different source bodies in the inner solar system, then also independently experienced this change in orbit, which eventually delivered them to Earth. It is also entirely possible that none of these rocks are related to each other and that each one represents a discrete source body within the solar system. Only as more orbital meteorites are collected and analyzed, can we more confidently infer about the nature of contemporary H chondrite source bodies.

**Table 12.** Other orbital meteorites that may be related to Arpu Kuilpu, based on orbital parameters and CRE age.

| Meteorite | a (AU) | i (deg) | e | Tj | CRE (Myr) | Ref. |
|---|---|---|---|---|---|---|
| Arpu Kuilpu (H5) | 2.75 ± 0.03 | 2.03 ± 0.01 | 0.671 ± 0.003 | 2.97 | 6-8 | a, this work |
| Košice (H5) | 2.71 ± 0.24 | 2 ± 0.8 | 0.647 ± 0.032 | 3.02 | 5-7 | b, c, |
| Morávka (H5) | 1.85 ± 0.07 | 32.2 ± 0.5 | 0.47 ± 0.02 | 3.70 | 5.7-7.7 | d, e |
| Lost City (H5) | 1.66 | 12 | 0.417 | 4.14 | 5.5 ± 0.5 | f, g, h |
| Hamburg (H4) | 2.73 ± 0.05 | 0.604 ± 0.11 | 0.661 ± 0.006 | 2.99 | 12 | i, j |
| Murrili (H5) | 2.52 ± 0.08 | 3.32 ± 0.06 | 0.609 ± 0.012 | 3.16 | 7 ± 1 | k, l |
| Ejby (H5/6) | 2.81 ± 0.09 | 0.96 ± 0.1 | 0.655 ± 0.011 | 2.96 | 83 ± 11 | m, n |

*a: Shober et al. (2022); b: Borovička et al. (2013); c: Povinec et al. (2015); d: Borovička et al. (2003a); e: Borovička et al. (2003b); f: Ceplecha and ReVelle, (2005); g: McCrosky et al. (1971); h: Bogard et al. (1971); i: Brown et al. (2019); j: Heck et al. (2020); k: Sansom et al. (2020); l: Anderson et al. (2021); m: Spurný et al. (2017); n: Haack et al. (2019);*

**CONCLUSIONS**

The Arpu Kuilpu orbital meteorite is a minimally-shocked, lightly-weathered, unbrecciated, H5 chondrite that was delivered to Earth on a JFC-like orbit. The very minor impact-related shock pressure it experienced occurred very early in the solar system's history. It was likely separated from its contemporary, asteroidal source body approximately 6-8 Ma ago, and may be related to other orbital H chondrites, with Košice being the closest match when considering prior orbits and CRE ages (see Discussion). Combining Arpu Kuilpu's asteroidal composition with its JFC-like orbit prior to impact further confirms the observation that small meteoroids (cm-m size) on JFC orbits can originate from the asteroid belt. Below (**Table 13**) we list the major features of this meteorite, and the methods used to determine them.

**Table 13.** The main features of the Arpu Kuilpu meteorite.

| Feature | Method(s) | Result |
|---|---|---|
| Petrologic Type | SEM-EDS, Optical Microscopy, EPMA | H5 |
| Meteorite Class | Gamma-Ray Spectrometry, Noble Gas Analysis, ICP-OES, ICP-MS, Mössbauer Spectroscopy, Ideal Gas Pycnometry, Magnetic Susceptibility, µCT | H |
| Shock State | Optical Microscopy | S1-S2 |
|  | µCT | S1-S2 |
| Weathering State | Visual Inspection, Mössbauer Spectroscopy | W0-W1 |
| Mineral Chemistry | EPMA | Olivine: Fa: $19.2 \pm 0.2$<br>Orthopyroxene: Fs: $16.8 \pm 0.2$<br>Orthopyroxene: Wo: $1.4 \pm 0.2$<br>Chromite: Cr/Cr+Al: $85.5 \pm 0.3$<br>Chromite: Fe/Fe+Mg: $84.4 \pm 1.2$ |
| Impact Chronology | $^{40}$Ar / $^{39}$Ar Chronology | $4467 \pm 16$ Ma |
| Magnetic Susceptibility | Susceptibility Meter | $\log \chi = 5.16 \pm 0.08$ |
| Bulk Density | µCT | 3.63 g cm$^{-3}$ |
|  | Ideal gas Pycnometry | $3.66 \pm 0.02$ g cm$^{-3}$ |
| Grain Density | Ideal gas Pycnometry | $3.77 \pm 0.02$ g cm$^{-3}$ |
| Porosity | Ideal gas Pycnometry | $(3.0 \pm 0.7)$ % |
| Oxygen Isotopes | Laser-assisted Fluorination | $\delta^{17}$O‰ = $2.904 \pm 0.177$;<br>$\delta^{18}$O‰ = $4.163 \pm 0.336$;<br>$\Delta^{17}$O‰ = $0.740 \pm 0.002$. |
| Meteoroid Size (radius) | Gamma-ray Spectrometry, AMS, Noble Gas Analysis | ~5 cm |
| Burial Depth in Meteoroid | Gamma-ray Spectrometry, AMS, Noble Gas Analysis | ~1 cm |
| Cosmic Ray Exposure Age | AMS, Noble Gas Analysis | 6-8 Myr |

**ACKNOWLEDGMENTS**

We would like to thank the Maralinga Tjarutja community for permission to search on their land, as well as providing the name for this meteorite. We also thank C. Alexander and J. Bridges for their careful and constructive reviews. This work was funded by the Australian Research Council's Discovery Project Scheme (DP170102529), and has been partially carried out within the framework of the Swiss NSF (NCCR PlanentS) under grant 51NF40 205606.

**REFERENCES**

Alaerts L., Lewis R. S., Anders, E. 1979. Isotopic anomalies of noble gases in meteorites and their origins—III. LL-chondrites. *Geochimica et Cosmochimica Acta* 43: 1399-1415. https://doi.org/https://doi.org/10.1016/0016-7037(79)90134-0

Alexander C. M. O'D. 2019. Quantitative models for the elemental and isotopic fractionations in the chondrites: The non-carbonaceous chondrites. *Geochimica et Cosmochimica Acta* 254: 246-276. https://doi.org/10.1016/j.gca.2019.01.026

Anders E. and Grevesse N. 1989. Abundances of the elements: Meteoritic and solar. *Geochimica et Cosmochimica Acta* 53: 197–214.

Anderson S. Benedix, G. K. Forman, L. V., Daly, L. Greenwood, R. C. Franchi, I. A. Friedrich, J. M. Macke, R., Wiggins, S., Britt D. Cadogan J. M. Meier M. M. M. Maden C. Busemann H. Welten, K. C., Caffee, M. W., Jourdan F. Mayers C. Kennedy T. Godel B. Esteban L. Merigot K. Bevan A. W. R. Bland P. A. Paxman J. Towner M. C. Cupak M. Sansom E. K. Howie R. Devillepoix H. A. R. Jansen-Sturgeon T. Stuart D. Strangway D. 2021. Mineralogy, petrology, geochemistry, and chronology of the Murrili (H5) meteorite fall: The third recovered fall from the Desert Fireball Network. *Meteoritics & Planetary Science* 56(2):241-259.

Anderson S. 2023. Farm to Table Meteorite: An End to End Exploration of the Solar System's Past, Present and Future. PhD Thesis, Curtin University, Perth. https://espace.curtin.edu.au/handle/20.500.11937/92481

Arpesella C. 1996. A low background counting facility at Laboratori Nazionali del Gran Sasso. Appl. Radiation Isotopes 47:991-996.

Bartol Neutron Monitors, 2017. http://neutronm.bartol.udel.edu/

Bhandari N. Bonino G. Callegari E. Cini Castagnoli G. Mathew K. J. Padia J. T. and Queirazza G. 1989. The Torino H6 meteorite shower. *Meteoritics* 24: 29–34.

Bhandari N. Mathew K.J. Rao M.N. Herpers U. Bremer K. Vogt S. Wölfli W. Hofmann H.J. Michel R. Bodemann R. Lange H.-J. 1993. Depth and size dependence of cosmogenic nuclide production rates in stony meteoroids. *Geochimica Cosmochimica Acta* 57: 2361-2375.

Bland P. A. Spurný P. Bevan A. W. R. Howard K. T. Towner M. C. Benedix G. K. Greenwood R. C. Shrbený L. Franchi I. A. Deacon G. Borovička J. 2012. The Australian Desert Fireball Network: A new era for planetary science. *Australian Journal of Earth Sciences* 59(2): 177-187.

Blott S. J. and Pye K. 2008. Particle Shape: A Review and New Methods of Characterization and Classification. *Sedimentology* 55: 31–63.

Bogard D. D. Clark R. S. Keith J. E. Reynolds M. A. 1971. Noble gases and radionuclides in Lost City and other recently fallen meteorites. *Journal of Geophysical Research* 76(17): 4076-4083.

Bonino G. Bhandari N. Murty S. V. S. Mahajan R. R. Suthar K. M. Shukla A. D. Shukla P. N. Cini Castagnoli G. and Taricco C. 2001. Solar and galactic cosmic ray records of the Fermo (H) chondrite regolith breccia. *Meteoritics & Planetary Science* 36: 831–839.

Borovička J. Spurný P. Kalenda P. and Tagliaferri E. 2003a. The Morávka meteorite fall: 1 Description of the events and determination of the fireball trajectory and orbit from video records. Meteoritics & Planetary Science 38:975-987.

Borovička J. Weber H. W. Jopek T. Jakeŝ P. Randa Z. Brown P. G. ReVelle D. O. Kalenda P. Schultz L. Kucera J. and Haloda J. 2003b. The Morávka meteorite fall: 3. Meteoroid initial size, history, structure, and composition. *Meteoritics & Planetary Science* 38(7):1005-102

Borovička J. Tóth J. Igaz A. Spurný P. Kalenda P. Haloda J. Svoreå J. Kornoš L. Silber E. Brown P. Husárik M. 2013. The Košice meteorite fall: Atmospheric trajectory, fragmentation, and orbit. *Meteoritics & Planetary Science* 48:1757-1779.

Bottke Jr W. F. Vokrouhlicky D. Brož M. Nesvorny D. 2001. Dynamical spreading of asteroid families by the Yarkovsky effect. *Science* 294(5547): 1693-1696.

Britt D. T. and Consolmagno G.J. 2003. Stony meteorite porosities and densities: A review of the data through 2001. *Meteoritics and Planetary Science* 38: 1161–1180.

Brown P. G. Vida D. Moser D. E. Granvik M. Koshak W. J. Chu D. Steckloff J. Licata A. Hariri S. Mason J. Mazur M. Cooke W. Krzeminski Z. 2019. The Hamburg Meteorite Fall: Fireball trajectory, orbit and dynamics. *Meteoritics & Planetary Science* 54(9):2027-2045.

Busemann H. Baur H. Wieler R. 2000. Primordial noble gases in "phase Q" in carbonaceous and ordinary chondrites studied by closed-system stepped etching. *Meteoritics and Planetary Science* 35: 949-973. https://doi.org/10.1111/j.1945-5100.2000.tb01485.x

Cassata W. S., Renne P. R. and Shuster D. L. (2011) Argon diffusion in pyroxenes: Implications for thermochronometry and mantle degassing. *Earth Planetary Science Letters* 304:407–416.

Ceplecha Z. and ReVelle D. O. 2005. Fragmentation model of meteoroid motion, mass loss, and radiation in the atmosphere. Meteoritics & Planetary Science 40:35-54.

Clayton R. N. Mayeda T. K. Goswami J. N. Olsen E. J. 1991. Oxygen isotope studies of ordinary chondrites. *Geochimica et Cosmochimica Acta* 55(8): 2317-2337.

Consolmagno G. J. Macke R. J. Rochette P. Britt D. T. Gattacceca J. 2006. Density, magnetic susceptibility, and the characterization of ordinary chondrite falls and showers. *Meteoritics & Planetary Science*, 41(3): 331-342.

Consolmagno G. J. Britt D. T. Macke R. J. 2008. The significance of meteorite porosity and density. *Chemie der Erde* 68: 1–29.

Devillepoix H. A. R. Cupák M. Bland P. A. Sansom E. K. Towner M. C. Howie R. M. Hartig B. A. D. Jansen-Sturgeon T. Shober P. M. Anderson S. L. Benedix G. K. Busan D. Sayers R. Jenniskens P. Albers J. Herd C. D. K. Hill P. J. A. Brown P. G. Krzeminski Z. Osinski G. R. Chennaoui Aoudjehane H. Benkhaldoun Z. Jabiri A. Guennoun M. Barka A. Darhmaoui H. Daly L. Collins G. S. McMullan S. Suttle M. D. Ireland T. Bonning G. Baeza L. Alrefay T. Y. Horner J. Swindle T. D. Hergenrother C. W. Fries M. D. Tomkins A. Langendam A. Rushmer T. O'Neill C. Janches D. Hormaechea J. L. Shaw C. Young J. S. Alexander M. Mardon A. D. Tate J. R. 2020. A global fireball observatory. *Planetary and Space Science* 191:105036.

Dyar D. M., Agresti D. G., Schaefer M. W., Grant C. A., Sklute E. C. 2006. Mössbauer spectroscopy of earth and planetary materials. *Annu. Rev. Earth Planet. Sci.*, 34(1):83-125

Eberhardt P. Geiss J. Lutz H. 1963. Neutrons in meteorites. In *Earth Science and Meteoritics*, edited by Geiss, J. and Goldberg, E. D. Amsterdam: North Holland Publ. Co. 143-168.

Fernández J. A. and Sosa A. 2015. Jupiter family comets in near-Earth orbits: are some of them interlopers from the asteroid belt? *Planetary and Space Science* 118:14-24.

Friedrich J. M. Wang M.-S. Lipschutz M. E. 2002. Comparison of the trace element composition of Tagish Lake with other primitive carbonaceous chondrites. *Meteoritics and Planetary Science* 37: 677-686.

Friedrich J. M. Wang M. S. Lipschutz M.E. 2003. Chemical studies of L chondrites. V: compositional patterns for 49 trace elements in 14 L4-6 and 7 LL4-6 falls. *Geochimica et Cosmochimica Acta* 67(13): 2467-2479.

Friedrich J. M. Wignarajah D. P. Chaudhary S. Rivers M. L. Nehru C. E. Ebel D. S. 2008. Three-dimensional petrography of metal phases in equilibrated L chondrites–Effects of shock loading and dynamic compaction. *Earth and Planetary Science Letters* 275: 172-180. doi:10.1016/j.epsl.2008.08.024

Friedrich J. M. Ruzicka A. Macke R. J. Thostenson J. O. Rudolph R. A. Rivers M. L. Ebel D. S. 2017. Relationships among physical properties as indicators of high temperature deformation or post-shock thermal annealing in ordinary chondrites. *Geochimica et Cosmochimica Acta* 203: 157-174. doi: 10.1016/j.gca.2016.12.039

Gattacceca J. Eisenlohr P. and Rochette P. 2004. Calibration of in situ magnetic susceptibility measurements. *Geophysical Journal International* 158:42-49.

Gattacceca J. Rochette P. Denise M. Consolmagno G. Folco L. 2005. An impact origin for the foliation of chondrites. *Earth and Planetary Science Letters* 234: 351-368.

Gattacceca J. McCubbin F. M. Grossman J. Bouvier A. Chabot N. L. d'Orazio M. Goodrich C. Greshake A. Gross J. Komatsu M. and Miao B. 2022. The Meteoritical Bulletin, No. 110. *Meteoritics & Planetary Science* 57(11):2102-2105.

Gattacceca J. McCubbin F. M. Grossman J. Schrader D. L. Cartier C. Consolmagno G. Goodrich C. Greshake A. Gross J. Joy K. Miao B. Zhang B. 2024. The Meteoritical Bulletin, No. 112. *Meteoritics and Planetary Science*

Chabot N. L. d'Orazio M. Goodrich C. Greshake A. Gross J. Komatsu M. and Miao B. 2022. The Meteoritical Bulletin, No. 110. *Meteoritics & Planetary Science* 57(11):2102-2105.

Greenwood R. C. Burbine T. H. Miller M. F. Franchi I. A. 2017. Melting and differentiation of early-formed asteroids: The perspective from high precision oxygen isotope studies. *Geochemistry* 77(1): 1-43.

Haack H. Sørensen A. N., Bischoff A., Patzek M., Barrat J.-A., Midtskogen S., Stempels E., Laubenstein M., Greewood R., Schmitt-Kopplin E., Busemann H., Maden C., Bauer K., Morino P., Schönbächler M., Voss P., Dahl-Jensen T. 2019. Ejby—A new H5/6 ordinary chondrite fall in Copenhagen, Denmark. *Meteoritics & Planetary Science* 54: 1853-1869.

Heck P. R. Greer J. Boesenberg J. S. Bouvier A. Caffee M. W. Cassata W. S. Corrigan C. Davis, A. M. Davis D. W. Fries M. Hankey M. Jenniskens P. Schmitt-Kopplin P. Sheu S. Trappitsch R. Velbel M. Weller B. Welten K. Yin Q-Z. Sanborn M. Ziegler K. Rowland D. Verosub K. Zhou Q. Liu Y. Tang G. Li Q. Li X. Zajacz Z. 2020. The fall, recovery, classification, and initial characterization of the Hamburg, Michigan H4 chondrite. *Meteoritics & planetary science* 55(11): 2341-2359.

Jelínek V. 1981. Characterization of the magnetic fabric of rocks. *Tectonophysics* 79: T63–T67.

Jenniskens P., Rubin A. E., Yin Q. Z., Sears D. W., Sandford S. A., Zolensky M. E., Krot A. N., Blair L., Kane D., Utas J., Verish R., Friedrich J. M., Wimpenny J., Eppich G. R., Ziegler K., Verosub K. L., Rowland D. J., Albers J., Gural P. S., Gigsby B., Fries M. D., Matson R., Johnston M., Silber E., Brown P., Yamakawa A., Sanborn M., Laubenstein M., Welten K. C., Nishiizumi K., Meier M. M. M., Busemann H., Clay P., Caffee M. W., Schmitt-Kopplin P., Hertkorn N., Glavin D. P., Callahan M. P., Dworkin J. P., Wu Q., Zare R. N., Grady M., Verchovsky S., Emel'Yanenko V., Naroenkov S., Clark D. L., Girten B., Worden P. S. 2014. Fall, recovery, and characterization of the Novato L6 chondrite breccia. *Meteoritics & Planetary Science* 49(8):1388-1425.

Jourdan F. Kennedy T. Benedix G. K. Eroglu E. Mayer C. 2020. Timing of the magmatic activity and upper crustal cooling of differentiated asteroid 4 Vesta. *Geochimica et Cosmochimica Acta* 273: 205-225.

Kallemeyn G. W. Rubin A. E. Wang D. and Wasson J. T. 1989. Ordinary chondrites – Bulk compositions, classification, lithophile-element fractionations, and composition-petrographic type relationships. *Geochimica et Cosmochimica Acta* 53: 2747-2767.

Kennedy T., Jourdan F., Eroglu E. and Mayers C. 2019. Bombardment history of asteroid 4 Vesta recorded by brecciated eucrites: Large impact event clusters at 4.50 Ga and discreet bombardment until 3.47 Ga. *Geochimica et Cosmochimica Acta* 260:99–123

Ketcham R. A. 2005. Computational Methods for Quantitative Analysis of Three-Dimensional Features in Geological Specimens. *Geosphere* 1: 32–41.

Kessel R. Beckett J. Stolper E. 2007. The thermal history of equilibrated ordinary chondrites and relationship between textural maturity and temperature. *Geochimica et Cosmochimica Acta* 71: 1855-1881. https://doi.org/10.1016/j.gca.2006.12.017

Kohman T. P. and Bender M. L. 1967. Nuclide production by cosmic rays in meteorites and on the Moon. High-Energy Nuclear Reactions in Astrophysics. In: High-Energy Nuclear Reactions in Astrophysics - A collection of articles. edited by Shen B. S. P. and Benjamin W. A. New York. N.Y.: 169–245.

Kohout T. Haloda J. Halodová J. Meier M. M. M. Maden C. Busemann H. Laubenstein M. Caffee M. W. Welten K. C. Hpp J. Trieloff M. Mahajan R. R. Naik S. Trigo-Rodríguez J. M. Moyano-Cambero C. E. Oshtrakh M. I. Maksimova A. A. Chukin A. V. Semionkin V. A. Karabanalov M. S. Felner I. Petrova E. V. Brusnitsyna E. V. Grokhovsky V. I. Yakovlev G. A. Gritsevich M. Lyytinen E. Moilanen J. Kruglikov N. A. Ishchenko A. V. 2017. Annama H chondrite—Mineralogy, physical properties, cosmic ray exposure, and parent body history. *Meteoritics & Planetary Science* 52: 1525-1541.

Laubenstein M. 2017. Screening of materials with high purity germanium detectors at the Laboratori Nazionali del Gran Sasso. *International Journal of Modern Physics A* 32(30):1743002.

Leya I. and Masarik J. 2009. Cosmogenic nuclides in stony meteorites revisited. *Meteoritics & Planetary Science* 44: 1061-1086. https://doi.org/10.1111/j.1945-5100.2009.tb00788.x

Lodders K. and Fegley B. 1998. The planetary scientist's companion / Katharina Lodders, Bruce Fegley. Oxford University Press.

Macke R. J. 2010. *Survey of Meteorite Physical Properties: Density, Porosity, and Magnetic Susceptibility.* Ph.D. Dissertation, University of Central Florida.

Macke R. J. Consolmagno G.J. Britt, D. T. 2011. Density, porosity, and magnetic susceptibility of carbonaceous chondrites. *Meteoritics & Planetary Science* 46(12): 1842-1862.

Macke R. J. Kent J. J. Kiefer W. S. and Britt D. T. 2015. 3D-Laser-Scanning Technique Applied to Bulk Density Measurements of Apollo Lunar Samples. *LPSC XLVI* #1716.

Mason B. and Graham A. L. 1970 Minor and Trace Elements in Meteoritic Minerals, *Smithsonian Contributions to the Earth Sciences* 3: 1-17.

McCrosky R. E. Posen A. Schwartz G. and Shao C.-Y. 1971. Lost City Meteorite – Its Recovery and a Comparison with Other Fireballs. Journal of Geophysical Research 76:4090-4180.

Meier M. M. M. 2017. Meteoriteorbits. info-tracking all known meteorites with photographic orbits. *48th Annual Lunar and Planetary Science Conference*. Abstract #1964.

Miller M. F. Franchi I. A. Sexton A. S. Pillinger C. T. 1999. High precision $\delta^{17}O$ isotope measurements of oxygen from silicates and other oxides: method and applications. *Rapid Communications in Mass Spectrometry* 13(13): 1211-1217.

Moniot R. K. 1980. Noble-gas-rich separates from ordinary chondrites. *Geochimica et Cosmochimica Acta* 44: 253-271. https://doi.org/https://doi.org/10.1016/0016-7037(80)90136-2

Morris R. V., Klingelhofer G., Bernhardt B., Schroder C., Rodionov D. S., de Souza Jr, P. A., Yen A., Gellert R., Evlanov E. N., Foh J., Kankeleit E. 2004. Mineralogy at Gusev crater from the Mossbauer spectrometer on the Spirit rover. *Science* 305(5685):833-836.

Murad E. and Cashion J. 2011. Mössbauer spectroscopy of environmental materials and their industrial utilization. *Springer Science & Business Media*

Nagai H. Honda M. Imamura M and Kobayashi K. 1993. Cosmogenic $^{10}Be$ and $^{26}Al$ in metal, carbon and silicate of meteorites. *Geochimica et Cosmochimica Acta* 57: 3705-3723.

Nesvorný D. Jenniskens P. Levison H.F. Bottke W. F. Vokrouhlický D. Gounelle M. 2010. Cometary origin of the zodiacal cloud and carbonaceous micrometeorites. Implications for hot debris disks. *The Astrophysical Journal* 713(2):816.

Nier A. O. 1950. A redetermination of the relative abundances of the isotopes of carbon, nitrogen, oxygen, argon and potassium. *Physical Review* 77: 789-793.

Nishiizumi K. 2004. Preparation of $^{26}Al$ AMS standards. *Nuclear Instruments and Methods in Physics Research*, B223–224: 388–392.

Nishiizumi K. Imamura M. Caffee M. W. Southon J. R. Finkel R. C. McAninch J. 2007. Absolute calibration of $^{10}Be$ AMS standards. *Nuclear Instruments and Methods in Physics Research*, B258: 403–413.

Oberst J. Molau S. Heinlein D. Gritzner C. Schindler M. Spurny P. Ceplecha Z. Rendtel J. Betlem H. 1998. The "European Fireball Network": current status and future prospects. *Meteoritics & Planetary Science* 33(1): 49-56.

Povinec P. P. Masarik J. Sykora I. Kováčik A. Beno J. Meier M. M. M. Wieler R. Laubenstein M. Porubčan V. 2015. Cosmogenic nuclides in the Kosice meteorite: Experimental investigations and Monte Carlo simulations. *Meteoritics & Planetary Science* 50:880-892.

Riebe M. E. I. Welten K. C. Meier M. M. M. Wieler R. Barth M. I. F. Ward D. Laubenstein M. Bischoff A. Caffee M. W. Nishiizumi K. Busemann H. 2017. Cosmic-ray exposure ages of six chondritic Almahata Sitta fragments. *Meteoritics & Planetary Science*, 52: 2353-2374. https://doi.org/10.1111/maps.12936

Roth A. S., Trappitsch R., Metzler K., Hofmann B. A., Leya I. 2017. Neon produced by solar cosmic rays in ordinary chondrites. *Meteoritics & Planetary science*, 52(6):1155-1172.

Sansom E. K. Bland P. A. Towner M. C. Devillepoix H. A. Cupak M. Howie R. M. Jansen-Sturgeon T. Cox M. A. Hartig B. A. Paxman J. P. Benedix G. 2020. Murrili meteorite's fall and recovery from Kati Thanda. *Meteoritics & Planetary Science* 55(9): 2157-2168.

Scott E. R. Taylor G. J. Keil K. 1986. Accretion, metamorphism, and brecciation of ordinary chondrites: Evidence from petrologic studies of meteorites from Roosevelt County, New Mexico. *Journal of Geophysical Research: Solid Earth*, 91(B13): E115-E123.

Sharma P. Kubik P. W. Fehn U. Gove G. E. Nishiizumi K. and Elmore D. 1990. Development of $^{36}$Cl standards for AMS. *Nuclear Instruments and Methods in Physics Research* B52: 410-415.

Sharma P. Bourgeois M. Elmore D. Granger D. Lipschutz M. E. Ma X. Miller T. Mueller K. Rickey F. Simms P. Vogt S. 2000. PRIME lab AMS performance, upgrades and research applications. *Nuclear Instruments and Methods in Physics Research* B172: 112–123.

Shober P. M. Jansen-Sturgeon T. Sansom E. K. Devillepoix H. A. Towner M. C. Bland P. A. Cupák M. Howie R. M. Hartig B. A. 2020a. Where did they come from, where did they go: Grazing fireballs. *The Astronomical Journal* 159(5):191.

Shober P. M. Jansen-Sturgeon T. Bland P. A. Devillepoix H. A. R. Sansom E. K. Towner M. C. Cupák M. Howie R. M. Hartig B. A. D. 2020b. Using atmospheric impact data to model meteoroid close encounters. *Monthly Notices of the Royal Astronomical Society* 498(4):5240-5250.

Shober P. M. Sansom E. K. Bland P. A. Devillepoix H. A. R. Towner M. C. Cupák M. Howie R. M. Hartig B. A. D. Anderson S. L. 2021. The Main Asteroid Belt: The Primary Source of Debris on Comet-like Orbits. *The Planetary Science Journal* 2(98).

Shober P. M. Devillepoix H. A. R. Sansom E. K. Towner M. C. Cupak M. Anderson S. L. Benedix G. Forman L. Bland P. A. Howie R. M. Hartig B. A. D. Laubenstein M. Cary F. Langendam A. 2022. Arpu Kuilpu: an H5 from the outer main belt. *Meteoritics & Planetary Science* 57(6):1146-1157.

Spergel M. S. Reedy R. C. Lazareth O. W. Levy P. W. Slatest L. A. 1986. Cosmogenic neutron-capture-produced nuclides in stony meteorites. 16$^{th}$ Proceedings of the Lunar & Planetary Science Conference. *Journal of Geophysical Research Supplement* 91: D483-D494.

Spurný P. Borovička J. Baumgarten G. Haack H. Heinlein D. Sørensen A.N. 2017. Atmospheric trajectory and heliocentric orbit of the Ejby meteorite fall in Denmark on February 6, 2016. Planetary and Space Science 147:192-198.

Steiger R. H. and Jäger E. 1977. Subcommission on geochronology: Convention on the use of decay constants in geo- and cosmochronology. *Earth Planetary Science Letters* 36: 359-362. https://doi.org/10.1016/0012-821X(77)90060-7

Stöffler D. Hamann C. Metzler K. 2018 Shock metamorphism of planetary silicate rocks and sediments: Proposal for an updated classification system. *Meteoritics and Planetary Science* 53: 5–49.

Tosi A. Zucolotto M. E., Andrade D. P., Winter O. C., Mourão D. C., Sfair R., Ziegler K., Perez P. D., Suarez S., Ornellas I. D. Zurita, M. 2023. The Santa Filomena meteorite shower: Trajectory, classification, and opaque phases as indicators of metamorphic conditions. *Meteoritics & Planetary Science, 58(5):*621-642.

Trigo-Rodríguez J. M. Llorca J. Castro-Tirado A. J. Ortiz J. L. Docobo J. A. Fabregat J. 2006. The Spanish fireball network. *Astronomy & Geophysics* 47(6): 6-26.

Van Schmus W. R. and Wood J. A. 1967. A chemical-petrologic classification for the chondritic meteorites. *Geochimica et Cosmochimica Acta* 31(5): 747-765.

Wasson J. T. and Kallemeyn G. W. 1988. Compositions of Chondrites. *Philosophical Transactions of the Royal Society of London. Series A, Mathematical and Physical Sciences*: 325, No. 1587, The Solar System: Chemistry as a Key to Its Origin, 535-544.

Welten K. C. Nishiizumi K. Masarik J. Caffee M. W. Jull A. J. T. Klandrud S. E. Wieler, R. 2001. Cosmic-ray exposure history of two Frontier Mountain H-chondrite showers from spallation and neutron-capture products. *Meteoritics & Planetary Science* 36(2):301-317.

Welten K. C. Caffee M. W. Leya I. Masarik J. Nishiizumi K. and Wieler R. 2003 Noble gases and cosmogenic radionuclides in the Gold Basin L4-chondrite shower: Thermal history, exposure history and pre-atmospheric size. *Meteoritics & Planetary Science* 38(1):157-173.

Welten K. C. Caffee M. W. Hillegonds D. J. McCoy T. J. Masarik J. Nishiizumi K. 2011. Cosmogenic radionuclides in L5 and LL5 chondrites from Queen Alexandra Range, Antarctica: Identification of a large L/LL5 chondrite shower with a preatmospheric mass of approximately 50,000 kg. *Meteoritics & Planetary Science* 46(2):177-196.

Wieler R. 2002. Cosmic-ray-produced noble gases in meteorites. *Reviews of Mineralogy and Geochemistry* 47: 125-170. https://doi.org/10.2138/rmg.2002.47.5

Wieler R. Huber L. Busemann H. Seiler S. Leya I. Maden C. Masarik J. Meier M. M. M. Nagao K. Trappitsch R. Irving A. J. 2016. Noble gases in 18 Martian meteorites and angrite Northwest Africa 7812—Exposure ages, trapped gases, and a re-evaluation of the evidence for solar cosmic ray-produced neon in shergottites and other achondrites. *Meteoritics & Planetary Science* 51: 407-428. https://doi.org/10.1111/maps.12600

Wolf S. F. Compton J. R. Gagnon C. J. 2012. Determination of 11 major and minor elements in chondritic meteorites by inductively coupled plasma mass spectrometry. *Talanta 100:* 276-281.

Woodcock N.H. 1977 Specification of fabric shapes using an eigenvalue method. *Geological Society of America Bulletin* 88: 1231–1236.

Woodcock N. H. and Naylor M. A. 1983 Randomness testing in three-dimensional orientation data. *Journal of Structural Geology* 5: 539–548.

Zingg T. 1935. Beitrag zur Schotteranalyse: Die Schotteranalyse und ihre Anwendung auf die Glattalschotter. *Schweizerische Mineralogische und Petrographische Mitteilungen* 15: 39–140.